\documentclass[fleqn,usenatbib]{mnras}

\usepackage{newtxtext,newtxmath}
\usepackage{graphicx}
\usepackage[T1]{fontenc}

\DeclareRobustCommand{\VAN}[3]{#2}
\let\VANthebibliography\thebibliography
\def\thebibliography{\DeclareRobustCommand{\VAN}[3]{##3}\VANthebibliography}

\usepackage{graphicx}	
\usepackage{amsmath}	

\usepackage{amssymb}	

\usepackage{algorithm}
\usepackage{algorithmic}

\DeclareMathOperator*{\argmin}{argmin}

\newcommand{\aref}[1]{\hyperref[#1]{Appendix~\ref*{#1}}}

\title[The rotation of planet-hosting stars]{The rotation of planet-hosting stars}

\author[Sibony Y. et al.]{Yves Sibony$^{1, 2}$\thanks{E-mail: yves.sibony@unige.ch}
Ravit Helled$^{2}$, 
Robert Feldmann$^{2}$\\
$^{1}$Geneva Observatory, University of Geneva, Chemin Pegasi 51, CH-1290 Versoix, Switzerland\\
$^{2}$Institute for Computational Science, University of Zurich, Winterthurerstrasse 190, CH-8057 Zurich, Switzerland
}

\date{Accepted XXX. Received YYY; in original form ZZZ}

\pubyear{2021}

\begin{document}
\label{firstpage}
\pagerange{\pageref{firstpage}--\pageref{lastpage}}
\maketitle

\begin{abstract}
Understanding the distribution of angular momentum  during the formation of planetary systems is a key topic in astrophysics. 
Data from the \emph{Kepler} and \emph{Gaia} missions allow to investigate whether stellar rotation is correlated with the presence of planets around Sun-like stars. 
Here, we perform a statistical analysis of the rotation period of 493 planet-hosting stars. These are matched to a control sample, without detected planets, with similar effective temperatures, masses, radii, metallicities, and ages.  
We find that planet-hosting stars rotate on average $1.63 \pm 0.40$ days slower. 
The difference in rotation is statistically significant both in samples including and not including planets confirmed by radial velocity follow-up observations. We also analyse the dependence of rotation distribution on various stellar and planetary properties. Our results could potentially be explained by planet detection biases depending on the rotation period of their host stars in both RV and transit methods. Alternatively, they could point to a physical link between the existence of planets and stellar rotation, emphasising the need to understand the role of angular momentum in the formation and evolution planetary systems.
\end{abstract}


\section{Introduction} \label{sec:intro}
The Sun contains more than 99\% of the mass of the solar system, but less than 1\% of its angular momentum \citep[e.g.,][]{Ray2012}.
In fact, most of the current angular momentum of the solar system is located in the gas giants, Jupiter and Saturn, due to their high masses and large orbital distances. The different distribution of angular momentum and mass is largely a consequence of angular momentum transport processes during the protoplanetary disk phase  \citep[e.g.,][]{2011ARA&A..49...67W}.


The discovery of thousands of planets around other stars provides the opportunity to characterise the properties of planetary systems in a statistically robust manner. Ground based observations and space missions such as CoRoT \citep{Baglin2003}, Kepler \citep{Borucki2010,Koch2010} and TESS \citep{Ricker2015}  have quantified the radii, masses, and orbits of many  exoplanets as well as the properties of their host stars. These new data make it  possible to explore whether the properties of planet-hosting stars differ from those without planets. A question of interest is whether the angular momentum of stars is correlated with the occurrence of (massive) planets, i.e., do stars that host (massive) planets rotate slower than those without planets?
\par

Theoretical works have not studied this question directly, but focused on rotational evolution during the pre-main sequence phase and on star-planet interactions.
For instance, tidal interactions have been suggested to result in spin-down of initially fast-rotating stars \citep[e.g.][]{Bolmont2016}, while magnetic and tidal effects may lead to planets migrating or even colliding with the star, affecting the stellar rotation \citep[e.g.][]{Ahuir2019}.

Observationally, this topic has been investigated and no clear trend has been found. 
For example, while \citet{Ceillier2015} found no effect of the presence of small planets on their host star's rotation, \citet{Alves2010} suggested that stars with planets exhibit a surplus of angular momentum compared to stars without planets. 
In contrast, \citet{Berget2010,Gonzalez2015} report that planet hosts tend to spin slower than similar stars without planets.
In addition to these studies, \citet{PazChinchon2015} confirm a trend between stellar angular momentum and stellar mass \citep[the Kraft relation,][]{Kraft1967}. Finally, \citet{Gurumath2019} report a dependence of planetary orbital angular momentum on planetary mass, and that this dependence differs between single and multiple planetary systems.


The \emph{Kepler} mission has detected more than 2400 transiting exoplanets around Sun-like stars, providing us with numerous planetary radii and orbital periods. Moreover, the Gaia Data Release 2 \citep{GAIA2016,Gaia2018,Arenou2018,Lindegren2018} has allowed the determination of stellar properties for a large number of stars with unprecedented precision. 
In this work we create a homogeneously derived catalog based on {\it Kepler} and {\it Gaia} data including various stellar \citep{Berger2020} and planetary (\href{https://exoplanetarchive.ipac.caltech.edu/}{NASA exoplanet archive}) properties, as well as stellar rotation periods \citep{McQuillan2013b,McQuillan2014} to investigate the  correlation between planet occurrence and stellar rotation. 
\par
Our paper is organised as follows. We detail our methods in \autoref{sec:methods}. Our results are presented in \autoref{sec:results}, where we first show the influence of the presence of a planet, and then investigate the dependence on stellar and planetary properties. We discuss our results and conclude in \autoref{sec:discussion}. 

\section{Methods} \label{sec:methods}

In this study, we combine rotation period measurements from \citet{McQuillan2013b}\footnote{Note that the stellar rotation is represented by a single value corresponding to uniform rotation while in reality stars can rotate differentially.} and \citet{McQuillan2014} with other fundamental stellar properties derived homogeneously by \citet{Berger2020}, to obtain a large dataset of 32,049 stars (493 with and 31,556 without detected planets). 
This dataset is large enough to infer statistically significant results. Furthermore, the homogeneously derived properties from \citet{Berger2020} make our datasets as self-consistent as possible, instead of combining data from many different sources. Measurement uncertainties are taken into account by bootstrapping the data. We then match the non-rotational properties of the two populations in order to remove biases related to the non-rotational stellar properties. This allows us to isolate a potential correlation between planet occurrence and stellar rotation. We also study the dependence of this correlation on stellar and planetary parameters.

We compare the rotation periods of stars with planets and stars without planets. The first step is to establish these two datasets: (i) stars with detected planets and (ii) stars without detected planets. More details can be found in \aref{Appendix1}. The stellar parameters we use are the stellar effective temperature $T_{\rm{eff}}$, mass $M_s$, radius $R_s$, metallicity [Fe/H], age $t_{\rm{age}}$, and rotation period $P_{\rm{rot}}$.\\

\subsection{Datasets} \label{subsec:datasets}

In order for the datasets to be self-consistent, we want as few parameter sources as possible. Rotation periods are obtained from \citet{McQuillan2013b,McQuillan2014}, which provide periods for 737 \emph{Kepler} Objects of Interest (KOI) and 34,030 \emph{Kepler} targets respectively. The methods used by the authors to determine the rotation periods are described in \citet{McQuillan2013a}.\\
We collect all other parameters from \citet{Berger2020}, where the  stellar properties for 186,301 \emph{Kepler} stars are determined  from isochrones and broadband photometry, Gaia Data Release 2 \citep{Arenou2018,GAIA2016,Gaia2018,Lindegren2018} parallaxes, and spectroscopic metallicities, where available \citep[for more details on their methods, see][]{Howes2019}. A ``goodness-of-fit" (GOF) parameter is supplied for each star and \citet{Berger2020} consider those with GOF\textless0.99 to have unreliable ages, and therefore we remove them from the datasets.\\
The planets' properties are taken from the \href{https://exoplanetarchive.ipac.caltech.edu/}{NASA exoplanet archive} (consulted on the 27th of April, 2020). Since we focus on transiting systems, the properties are the planetary radius $R_p$ and orbital period $P_{\rm{orb}}$, as well as the number of planets detected in each system (multiplicity).\\
We cross-reference the two stellar catalogues by KIC identification number. We separate the stars from \citet{McQuillan2013b} (KOIs) that appear in the \href{https://exoplanetarchive.ipac.caltech.edu/}{NASA exoplanet archive} to make the set of stars with planets, and the rest of the stars make up the control sample. The final dataset contains 32,049 stars, split into 493 stars with and 31,556 stars without confirmed planets. \autoref{table:spectral_types} shows the distribution of spectral types of the planet-hosting stars in the dataset.
\begin{figure*}
    \centering
    \includegraphics[scale=.9]{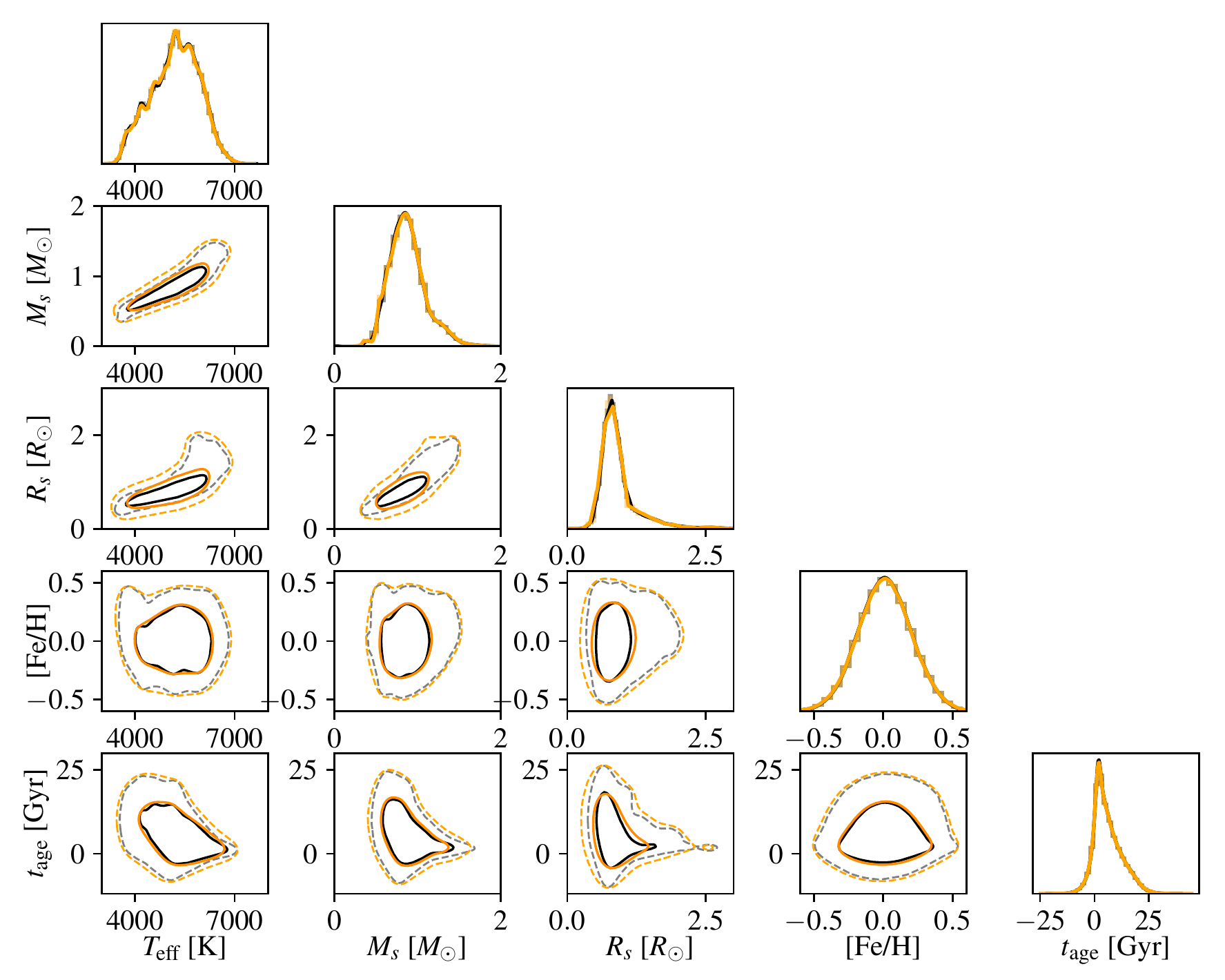}
    \caption{Distributions of the various non-rotational stellar properties both for the bootstrapped planet-hosting stars and for the matched control sample. Diagonal plots show one-dimensional probability distribution functions of the two data sets, while off-diagonal plots show two-dimensional distributions. In each panel, orange (black) lines and points represent  stars with (without) detected planets. Contour lines enclose 68\% (solid) and 95\% (dashed) of all the stars.}
    \label{fig:pairplots}
\end{figure*}
\begin{table}
\caption{Number of planet-hosting stars belonging to each spectral type.} \label{table:spectral_types}
\begin{tabular}{ccc}
\hline
Spectral type & \hspace{45pt}$T_{\rm{eff}}$\hspace{45pt} & Number of stars\\
\hline
F & 6000 - 7500 & 68\\
G & 5000 - 6000 & 244\\
K & 3500 - 5000 & 180\\
M & 2500 - 3500 & 1\\
\hline
\end{tabular}
\end{table}
\subsection{Bootstrapping and matching processes} \label{subsec:matching}

The subset of stars with confirmed planets is not a representative sample of the total stellar population. Indeed, both physical effects and observational biases lead to differences in those populations' properties. We can evoke for example the widely accepted correlation between giant planet occurrence and stellar metallicity \citep{Gonzalez1997,Santos2004,Fischer2005,Mortier2013}, as well as the bias of the transit method towards stars of smaller radii. Our aim is to investigate whether, all else being equal, planet occurrence is correlated with stellar rotation. Thus, in order to remove those effects and compare apples to apples, we match each planet-hosting star with the most similar star in the control sample. By its nature, our matching procedure can only correct for biases of the non-rotational stellar properties but not those linked to that of the stellar rotation.\\

The stellar rotation periods are provided with $1\sigma$ errors by \citet{McQuillan2013b,McQuillan2014}, and the average relative error of the full cross-referenced sample is $\frac{\sigma(P_{\rm{rot}})}{P_{\rm{rot}}}=2.7\%$. The errors on the other stellar parameters ($T_{\rm{eff}}$, $M_s$, $R_s$, [Fe/H], and  $t_{\rm{age}}$) are given as upper and lower values by \citet{Berger2020}, and we keep the larger one of the two as the $1\sigma$ uncertainty. The resulting average relative errors on non-rotational parameters are:  
\begin{align*}
\hspace{30pt}
\frac{\sigma(T_{\rm{eff}})}{T_{\rm{eff}}}=1.9\%,
~\frac{\sigma(M_s)}{M_s}&=5.9\%,~ \frac{\sigma(R_s)}{R_s}=4.2\%,\\
 \frac{\sigma(\text{[Fe/H]})}{\text{[Fe/H]}}=34.1\%&,~\text{and}~ \frac{\sigma(t_{\rm{age}})}{t_{\rm{age}}}=78.5\%.
\end{align*}

First, we bootstrap both datasets within the uncertainty ranges of each stellar parameter. We sample every star (both with and without planets) 1000 times assuming a normal distribution on each parameter, with the standard deviation corresponding to the measurement uncertainties. We thus end up with 1000 samples of the two datasets. For every one of these 1000 samples, we use a multidimensional Euclidean distance based on the non-rotational parameters to match each planet-hosting star to the most similar star without detected planets. We then compare the rotation periods between the stars with and without detected planets. A detailed description of the methods is given in \aref{Appendix1}. 

\autoref{fig:pairplots} shows two-dimensional distributions of the various non-rotational stellar properties both for the planet-hosting stars and for the matched control sample showing excellent agreement. In addition, the 1-dimensional distributions of the non-rotational parameters agree well between the two samples, see \autoref{table1}, with $p$-values of Kolmogorov-Smirnov tests close to unity. A similar figure for the unmatched datasets can be found in \autoref{fig:pairplots_prematch}. Note, however, that contour lines in \autoref{fig:pairplots} include the bootstrapping of the measurement errors, while the contour lines in \autoref{fig:pairplots_prematch} do not.

\begin{figure}
    \centering
    \includegraphics[scale=1]{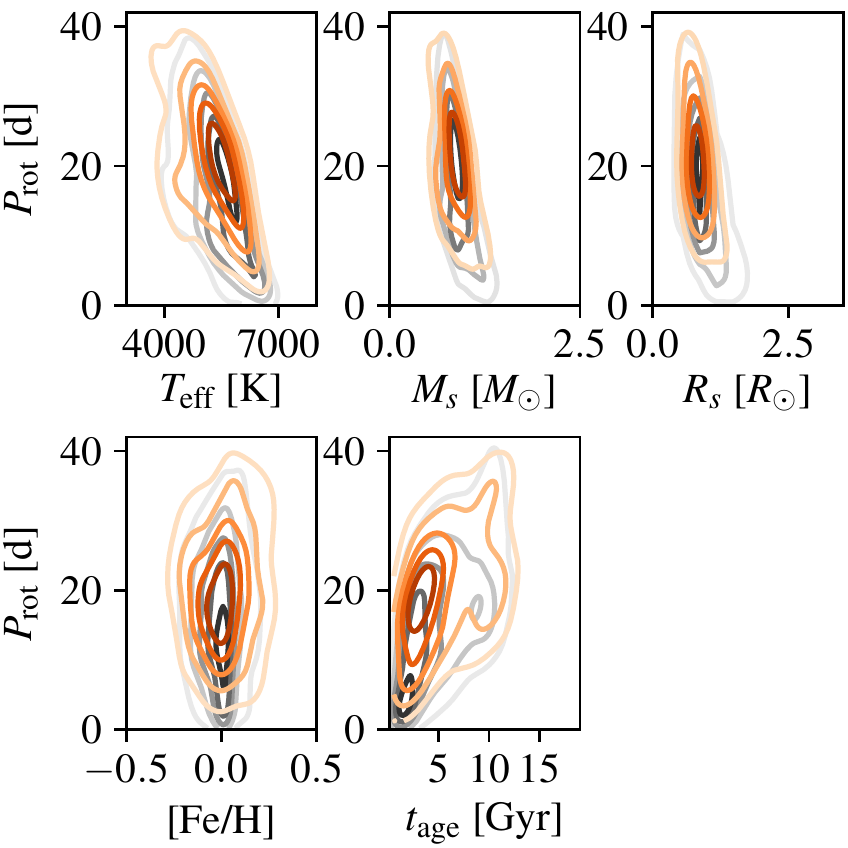}
    \caption{Bivariate contour plots of $P_{\rm rot}$ as a function of the non-rotational stellar properties. Orange (black) curves correspond to the final datasets of stars with (without) planets, see \autoref{subsec:datasets}.}
    \label{fig:nonrot_vs_Prot}
\end{figure}

\autoref{fig:nonrot_vs_Prot} shows $P_{\rm rot}$ as a function of the non-rotational stellar properties. 
The two populations correspond to stars with (orange) and without (black) planets (unmatched). It is clear that regardless to the existence of planets there are correlations between $P_{\rm rot}$ and the other stellar properties. Most significant are the higher $P_{\rm rot}$ for cooler, smaller and low-mass stars \citep[investigated by e.g.][]{McQuillan2014}. Also younger stars seem to rotate faster \citep[this is coherent with][]{Kraft1967,Skumanich1972}. 

\begin{table*}
\caption{Pearson and Spearman correlation coefficients between $\log{P_{\rm{rot}}}$ and the non-rotational stellar parameters.} \label{table:corr_coeffs}
\begin{tabular}{lccccc}
\hline
 & $\log{T_{\rm{eff}}}$ & $\log{M_s}$ & $\log{R_s}$ & [Fe/H] & $\log{t_{\rm{age}}}$\\
\hline
Pearson with planets  & $-0.501\pm0.039$ & $-0.496\pm0.039$ & $-0.438\pm0.041$ & $0.0516\pm0.0451$ &  $0.476\pm0.040$ \\
Pearson without planets  & $-0.485\pm0.00$5 & $-0.488\pm0.00$5 &  $-0.433\pm0.005$ &   $0.0321\pm0.006$ &  $0.378\pm0.005$ \\
Spearman with planets & $-0.518\pm0.029$ & $-0.481\pm0.029$ & $-0.443\pm0.029$ &   $0.086\pm0.029$ &   $0.456\pm0.029$ \\
Spearman without planets  & $-0.559\pm0.003$ & $-0.532\pm0.003$ & $-0.501\pm0.003$ &  $0.0702\pm0.003$ &  $0.393\pm0.003$ \\
\hline
\end{tabular}\\
{\raggedright
The correlation coefficients for the different stellar properties are given for both datasets of stars. \par}
\end{table*}

Our quantity of interest is the average $P_{\rm{rot}}$ difference between stars without planets and planet-hosting stars, $\Delta P_{\rm{rot}}$. It is obtained in the following manner: in each bootstrap iteration {\it i} we compute the average of all 493 pairwise differences, $\Delta P_{\rm{rot}}^i$. Then the mean and standard deviation over the 1000 $\Delta P_{\rm{rot}}^i$ are estimators of the overall average period difference $\Delta P_{\rm{rot}}$ and its error $e(\Delta P_{\rm{rot}})$, respectively.\\

In order to validate our method, we apply it to a random sample of stars without detected planets, biased in non-rotational parameters. By construction, the selection bias does not depend on stellar rotation, and therefore our method should return a $\Delta P_{\rm rot}$ value that is not statistically significant. This is indeed the result where $\Delta P_{\rm rot}=(0.38\pm0.35)\text{d}$. Further details on this validation can be found in \aref{Appendix2}.\\
Furthermore, a large fraction of Kepler planets have been confirmed with radial velocity (RV) follow-up observations. This could lead to the sample of stars with confirmed planets being biased in stellar rotation. Indeed a radial velocity detection is harder to obtain around a fast-rotating star because of the Doppler broadening induced by stellar rotation. To ensure that this effect is not the cause of the $\Delta P_{\rm rot}$ we detect, we remove from our datasets the stars whose planets have been confirmed by RV (see \aref{Appendix3} for further information). 

\begin{table*}
\caption{Statistical tests on matched datasets.} \label{table1}
\begin{tabular}{lcccccc}
\hline
 & $T_{\rm{eff}}$ & $M_s$ & $R_s$ & [Fe/H] & $t_{\rm{age}}$ & $P_{\rm{rot}}$\\
\hline
KS p-value&0.999&0.994&0.950&0.999&0.999&0.00480\\
KS statistic&0.0214&0.0238&0.0308&0.0195&0.0199&0.123\\
TT p-value&0.935&0.943&0.934&0.965&0.961&0.0183\\
TT statistic&-0.0726&0.0251&-0.0229&0.00390&0.00939&2.75\\
Wilcoxon p-value&0.416&0.479&0.434&0.475&0.503&0.00613\\
Wilcoxon statistic&57674&58194&57812&58200&58371&50039\\
\hline
\end{tabular}\\
{\raggedright Comparison of the distribution of stellar parameters with a Kolmogorov-Smirnov test (first two rows), a Student $t$ test (middle rows), and a Wilcoxon signed-rank test (bottom two rows) for stars with planets and for stars without planets after matching on non-rotational stellar properties. We perform each test for each of the 1000 bootstrap iterations of the matching, and then report the mean statistics. \par}
\end{table*}

\section{Results} \label{sec:results}
\subsection{Influence of the presence of a planet} \label{subsec:planetpresence}
\begin{figure*}
    \centering
    \includegraphics{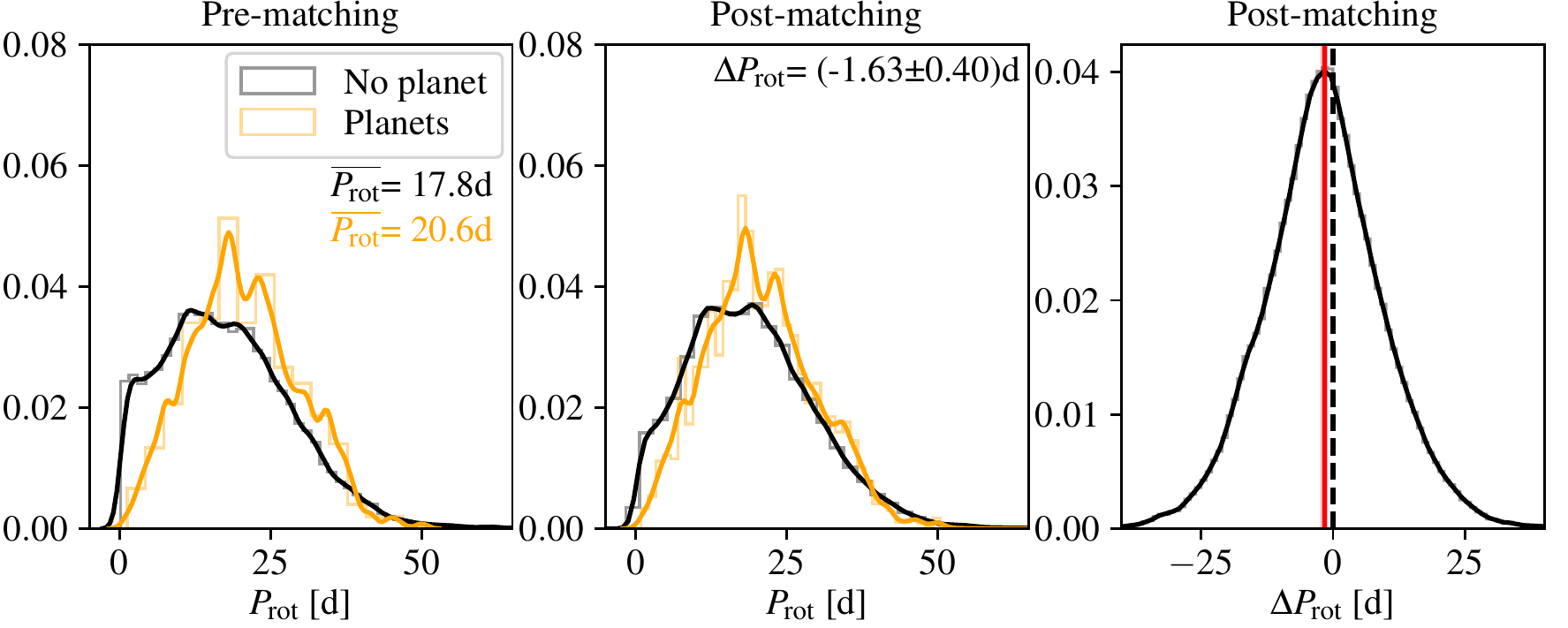}
    \caption{Left panel: Rotation period histograms for stars with one or more detected planets (orange) and stars without detected planets (black), before matching. We give the average rotation period $\overline{P_{\rm rot}}$ of each dataset. Middle panel: Rotation period histograms for stars with planets (orange) and their matched counterparts without detected planets (black). Also shown is the inferred $\Delta P_{\rm{rot}}$ (see  \aref{Appendix1} for details). Right panel: Histogram of $\Delta P_{\rm{rot}}$ for all 493000 matched pairs of (star with planets, star without planets). The average difference over all samples and iterations is $\Delta P_{\rm{rot}}= (-1.63\pm0.40)$d (indicated by the red line). The black dashed line corresponds to  $\Delta P_{\rm{rot}}=0$.}
    \label{fig:planetpresence}
\end{figure*}

The main question we aim to answer in this work is whether there is a statistically significant period difference between stars with and without planets.
The left panel of \autoref{fig:planetpresence}  shows the distribution of rotation periods of stars with planets (orange) and without detected planets (black). We indicate the average rotation period of each dataset; before matching the average period difference is -3.2 days.

The middle panel of \autoref{fig:planetpresence} shows the distributions of rotation periods for stars with and without planets, after matching.  We show the data for the entire bootstrapped sample of $1000\times493$ stars with planets and their matched counterparts.
After matching, we find a $4\sigma$ result:
\begin{equation}
    \Delta P_{\rm{rot}} = (-1.63\pm0.40)~\text{days}.
\end{equation}

This shows that the stars with planets rotate on average 1.6 days slower than those without, all else being equal.  We clearly see that, although the two matched populations are very similar in their non-rotational properties (see \autoref{fig:pairplots}), the rotation period distributions are very different. The p-values for $P_{\rm{rot}}$, shown in \autoref{table1} are $4.8\times10^{-3}$ for the KS test and $1.8 \times10^{-2}$ for the t-test.
The right panel of \autoref{fig:planetpresence} shows the distribution of period differences among matched pairs of (star with planets, star without planets) counterparts. The red line represents the result $\Delta P_{\rm{rot}} = (-1.63\pm0.40)~ \text{days}$, and the black dashed line indicates $\Delta P_{\rm{rot}} = 0$. 

The reported metallicities of many targets in our dataset are not based on spectroscopic measurements and therefore their values are somewhat uncertain. In order to check the importance of the metallicity values on our results, we  have cross-matched our catalogue with the LAMOST catalogue \citep[][\href{http://dr7.lamost.org/v2.0/}{LAMOST DR7}]{LAMOST2012}, yielding 133 stars with and 6075 stars without detected planets. Applying our method on these two datasets, we find $\Delta P_{\rm rot} = (-1.58 \pm 0.70)~$days, in very good agreement with the result obtained using all stellar properties from \citet{Berger2020}.\\
Another potential concern relates to stellar ages. While \citet{Berger2020} provide a GOF parameter for stellar ages, we still have stars in the dataset with estimated ages of $\gtrsim{}14$ Gyr, as well as stars with relative age uncertainties larger than 100\%. Therefore, we perform two different cuts on our datasets. First, we select only the stars with $t_{\rm age}<14$ Gyr and with relative uncertainties smaller than 100\%. This yields 264 stars with and 18837 stars without detected planets. For this sample we find $\Delta P_{\rm rot} = (-1.43 \pm 0.56)~$days. In a second, more stringent, test we consider only stars with relative age uncertainties of less than 50\%. This yields 57 stars with and 7394 stars without detected planets. For this data set we get $\Delta P_{\rm rot} = (-3.00 \pm 1.19)~$days. Both of these results, while less statistically significant, are still in agreement (within 3$\sigma$) with the result we get using the full  datasets.\\
In a further test, we artificially inflate the uncertainties on the stellar rotation periods by a factor of three finding again no significant change in our results. 
The uncertainties on stellar rotation do not come into play in the matching process (as we match on non-rotational properties), but only in the bootstrapping. This means that changing the uncertainties on stellar rotation does not change which stars get matched to one another. The individual differences in rotation periods between pairs of matched counterparts are changed by inflating the uncertainties, but they average out due to the large number of bootstraps.
\subsection{Dependence on stellar properties}
\label{subsec:stellar_param}

\begin{figure*}
    \centering
    \includegraphics{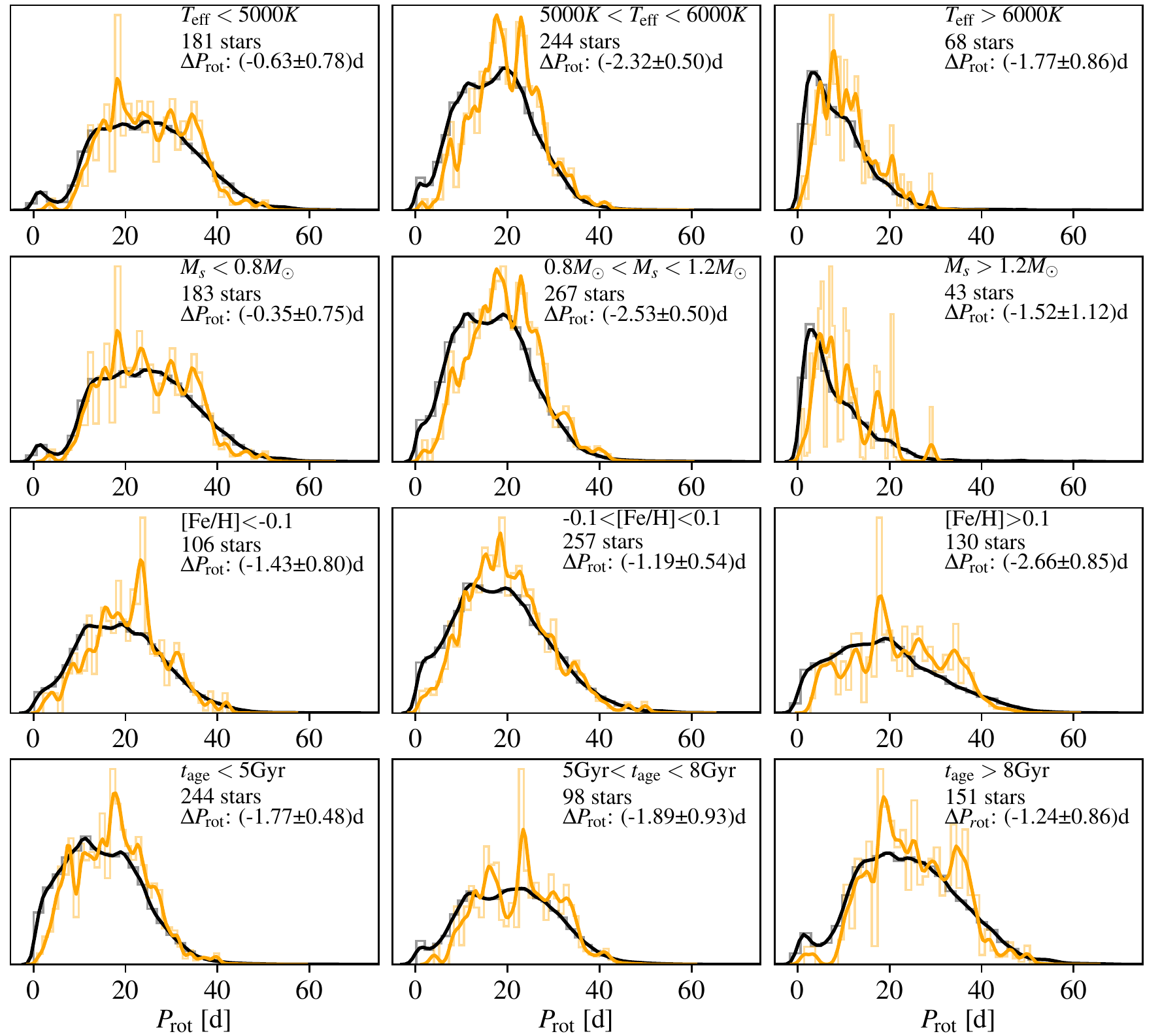}
    \caption{Rotation period histograms (after matching) for stars with and without detected planets in bins of stellar mass, effective temperature, metallicity, and age (from top to bottom).
    As in previous figures, the orange curves represent stars with planets and the black curves, stars without planets. In each panel we specify the selection criterion, the amount of stars, and the average rotation period difference.}
    \label{fig:stellar_param_Prots}
\end{figure*}

\begin{table*}
\caption{$\Delta P_{\rm{rot}}$ as a function of stellar parameters.} \label{table:stellar_param}
\begin{tabular}{lccccc}
\hline
Filter & $N_{\rm{stars}}$ & $\overline{P_{\rm{rot}}}(\text{with planets})$ [d] & $\overline{P_{\rm{rot}}}(\text{without planets})$ [d] & $\Delta P_{\rm{rot}}$ [d] & p-value\\
\hline
$T_{\rm{eff}}$ < 5000K & 181 & 25.2 & 24.6 & $-0.63 \pm 0.78$ & 0.37 \\
5000K < $T_{\rm{eff}}$ < 6000K & 244 & 19.9 & 17.6 & $-2.34 \pm 0.50$ & $1.4\times10^{-5}$ \\
$T_{\rm{eff}}$ > 6000K & 68 & 11.1 & 9.3 & $-1.77 \pm 0.86$ & 0.056 \\
$M_s$ < $0.8M_\odot$ & 183  & 24.8 & 24.4 & $-0.35 \pm 0.75$ & 0.48\\
$0.8M_\odot$ < $M_s$ < $1.2M_\odot$ & 267 & 19.5 & 16.9 & $-2.53 \pm 0.50$ & $2.2\times10^{-6}$\\
$M_s$ > $1.2M_\odot$ & 43 & 10.4 & 8.8 & $-1.52 \pm 1.12$ & 0.075\\
\text{[Fe/H]} < -0.1 & 106 & 20.6 & 19.2 & $-1.43 \pm 0.80$ & 0.10\\
-0.1 < [Fe/H] < 0.1 & 257 & 20.0 & 18.8 & $-1.19 \pm 0.54$ & 0.065\\
\text{[Fe/H]} > 0.1 & 130 & 21.9 & 19.3 & $-2.66 \pm 0.85$ & $3.5 \times 10^{-3}$\\
$t_{\rm{age}}$ < 5Gyr & 244 & 17.0 & 15.3 & $-1.77 \pm 0.48$ & $9.3\times10^{-4}$\\
5Gyr < $t_{\rm{age}}$ < 8Gyr & 98 & 22.6 & 20.7 & $-1.89 \pm 0.93$ & 0.054\\
$t_{\rm{age}}$ > 8Gyr & 151 & 25.2 & 24.0 & $-1.24 \pm 0.86$ & 0.17\\
\hline
\end{tabular}
\end{table*}
We next analyse the rotation period of the stars in the dataset depending on the stellar properties. Specifically, we divide the sample as follows. 
In terms of mass we split into low-mass ($M_s < 0.8M_\odot$), Sun-like mass ($0.8M_\odot < M_s < 1.2M_\odot$), and high-mass ($M_s > 1.2M_\odot$) stars.  We also separate the sample according to temperature: Cool ($T_{\rm{eff}}<5000K$), Sun-like effective temperature ($5000K < T_{\rm{eff}} < 6000K$), and Hot ($T_{\rm{eff}} > 6000K$), metallicity: low ([Fe/H]$<-0.1$),  intermediate  ($-0.1<$[Fe/H]$<0.1$), and high ([Fe/H]$>0.1$). Finally, we consider the stellar age: young ($t_{\rm{age}}<5$Gyr),  intermediate ($5\text{Gyr}<t_{\rm{age}}<8$Gyr), and old ($t_{\rm{age}}>8$Gyr). The distributions of rotation period of the various sub-populations are shown in \autoref{fig:stellar_param_Prots}. Interestingly, the average rotation period of massive and hot stars is $\sim$10 days, much shorter than the mean rotation period of the full dataset \citep[this is consistent with][]{Kraft1967,Skumanich1972, McQuillan2014}. In the remainder of this section we focus, however, on the difference in rotation period between stars with and without detected planets.

Intermediate-mass stars show a significantly smaller, i.e., more negative, $\Delta P_{\rm rot}\sim{}-2.5\pm{}0.5$d than the entire sample. In other words, stars with mass similar to the Sun do not only rotate slower than similar stars without detected planets but also slower (on average) than low-mass or high-mass stars with detected planets. The same results ($\Delta P_{\rm rot}\sim{}-2.3\pm{}0.5$d) also hold when we split our sample according to effective temperature instead of mass.

As can be seen in the third row of \autoref{fig:stellar_param_Prots} the $P_{\rm rot}$ distributions shifting to higher values (slower rotation) for higher metallicities. 
This result is consistent with previous studies suggesting that metal-rich stars rotate more slowly than metal-poor stars \citep[e.g.][]{Karoff2018, Amard2020}. We also find that metal-rich stars with planets rotate significantly slower than those without ($\Delta P_{\rm rot}\sim{}-2.7\pm{}0.9$d), while intermediate-metallicity stars show a weaker  (2-$\sigma$) difference in their rotation periods. Similarly, young stars with planets rotate much slower than young stars without detected planets ($\Delta P_{\rm rot}\sim{}-1.8\pm{}0.5$d). The difference in rotation is also noticeable for intermediate-age stars albeit with a lower statistical significance.
The other studied sub-populations do not show strong differences in their average rotation periods. Our findings are summarised in \autoref{table:stellar_param}. 

\begin{figure}
    \centering
    \hspace*{-30pt}
    \includegraphics[scale=.5]{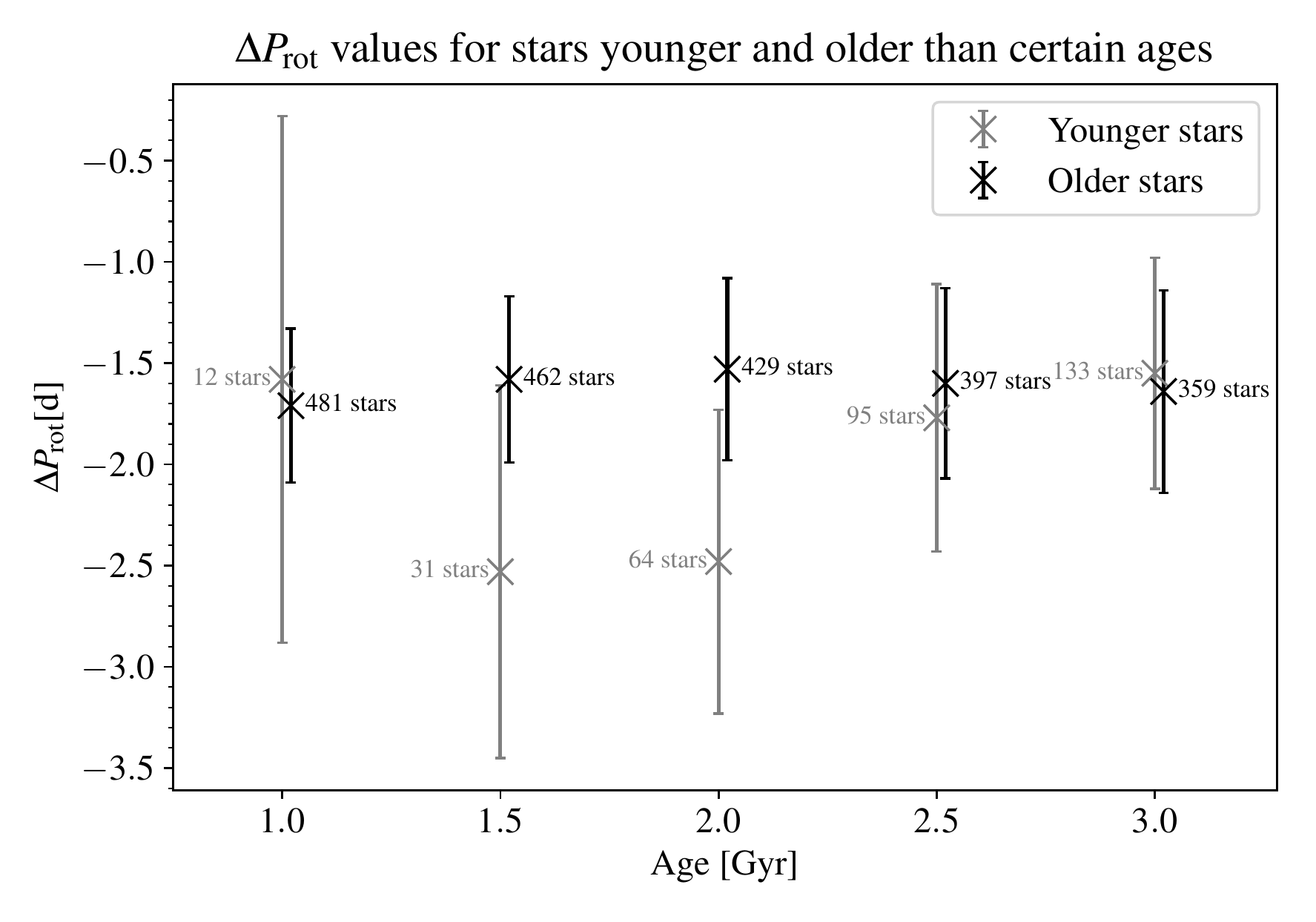}
    \caption{Scatter plot of the $\Delta P_{\rm rot}$ values for stars older and younger than certain ages. The values and uncertainties of $\Delta P_{\rm rot}$ and the corresponding  number of stars in each age bin are listed in \autoref{table:deltas_over_under}.}
    \label{fig:deltas_over_under}
\end{figure}

While theoretical models predict that magnetic braking slows down the stellar rotation within a Gyr timescale \citep[e.g.][]{Bouvier2008}, recent studies have extended this upper limit due to evidence of older stars having unusually fast rotation \citep{vanSaders2016,Hall2021}. We therefore analyse $\Delta P_{\rm rot}$ for stars with different ages. \autoref{fig:deltas_over_under} shows $\Delta P_{\rm rot}$ for stars below and above a given age limit ranging from 1 to 3 Gyr. As expected, the magnitude of $\Delta P_{\rm rot}$ decreases for samples with older stars.  Interestingly, however, even the sample consisting of stars with ages $>$ 3 Gyr still shows a statistically significant $\Delta P_{\rm rot}$. This observation can be understood from the scaling of the rotation period with stellar age shown in \autoref{fig:Age_vrot_All} in the Appendix. 
While the stellar rotation period increases quickly with age for stars $<1$ Gyr, the slow down in rotation continues up to several Gyr. We note that the relation between stellar age and rotation period shows a large scatter.

Finally, we also separate the sample according to the stellar rotation periods, selecting stars with and without detected planets that have $P_{\rm rot}>2~$days, $P_{\rm rot}>7~$days, $P_{\rm rot}>15~$days, and $P_{\rm rot}<15~$days. Removing  fast rotators (keeping stars with $P_{\rm rot}>2-7~$days) reduces the statistical significant of our result to 2-3$\sigma$. At the same time, when we consider only the stars with $P_{\rm rot}>15~$days we find that there is no difference in rotation period. This seems to suggest that our result is strongly affected by the fast-rotating stars in our dataset. 
There are however two points that make such a selection less than ideal. First, selecting stars according to their rotation period hinders the matching process. Indeed, for our analysis we match stars  based on their non-rotational properties. It is then possible for a slow-rotator to be best matched with a fast-rotator, or vice-versa. If however we perform a selection based on rotation period for both stars with and without planets, it is impossible for a slow-rotator to be matched with a fast-rotator, or vice-versa. The effect would be to remove the tails of the distribution of $\Delta P_{\rm rot}$ that we see in the right panel of \autoref{fig:planetpresence}. In other words, the best possible match for a certain star may be removed from the dataset due to the $P_{\rm rot}$ selection process, which leads to a rather large loss of information. Second, $P_{\rm rot}$ is correlated to other stellar properties \citep[see \autoref{fig:nonrot_vs_Prot} or e.g.][]{McQuillan2014}. Selecting only stars with $P_{\rm rot}>15~$days biases the samples towards low-mass, colder stars, which exhibit the least significant result, as shown in \autoref{fig:stellar_param_Prots}. It is therefore not surprising that our result does not hold for stars with $P_{\rm rot}>15~$days.
\subsection{Dependence on planetary properties}
\label{subsec:planetary_param}
In this section, we investigate how the rotation periods of stars  depend on the properties of their planets. We split our data based on planet multiplicity, orbital period $P_{\rm{orb}}$, and planet radius $R_p$ arriving at the following sub-samples.
Planetary systems with a `distant planet' correspond to systems with at least one planet with $P_{\rm{orb}}>10$d, while systems with `no distant planet' host only planets with $P_{\rm{orb}}<10$d. Similarly, systems with a `large planet'  have at least one planet with $R_p>0.3R_J$. In contrast, systems containing `no large planet' are those without a planet larger than $0.3R_J$.
`Single' and `multiple' planet systems are systems with one and more than one detected planets, respectively.
In addition, we define `Hot Jupiters' as systems having one or more planets with $P_{\rm{orb}}<10$d and $R_p>0.3R_J$, while `Cool Jupiters' have $P_{\rm{orb}}>10$d. Finally, `One close planet' and `One distant planet' are single-planet systems with $P_{\rm{orb}}<10$d and $P_{\rm{orb}}>10$d, respectively. The specific selection criteria of the various sub-samples are listed in \autoref{table:planetary_param}. We analyse the sensitivity of our results on our sample selection criteria in Appendix \autoref{table:appendix_table}.

All of the considered sub-samples show slower average stellar rotation than stars without planets. The statistical significance of this finding varies among the sub-samples due to the large spread in sub-sample sizes ranging from 37 to 412 stars. A slower average stellar rotation is naturally expected given that stars with planets rotate on average slower than stars without planets. The most significant difference in rotation period are found for systems without a large planet, single planet systems, and systems with a distant planet, see \autoref{table:planetary_param}. Other sub-samples show less significant results possibly due to their relatively small sample sizes.

\begin{table*}
\caption{Difference in rotation period of stars with planets depending on their planetary properties.} \label{table:planetary_param}
\begin{tabular}{lclll}
\hline
Name & Criterion & $N_{\rm{stars}}$ & $\Delta P_{\rm{rot}}$ [d] & p-value\\
\hline
Distant planet & $P_{\rm{orb}} > 10$d & 265 & -1.82 $\pm$ 0.53 & 0.0021\\
No distant planet & $P_{\rm{orb}} < 10$d & 228 & -1.41 $\pm$ 0.61 & 0.045\\
Large planet & $R_{\rm p} > 0.3R_{\rm J}$ & 81 & -1.59 $\pm$ 0.89 & 0.092\\
No large planet & $R_{\rm p} < 0.3R_{\rm J}$ & 412 & -1.64 $\pm$ 0.45 & 0.0013\\
Single planet & \text{Multiplicity} = 1 & 334 & -1.78 $\pm$ 0.5 & 0.0014\\
Multiple planets & \text{Multiplicity} $>$ 1& 159 & -1.32 $\pm$ 0.68 & 0.091\\
Hot Jupiter & $R_{\rm p} > 0.3$ $R_{\rm J}$ ~\&~ $P_{\rm{orb}} < 10$d & 37 & -1.27 $\pm$ 1.37 & 0.19\\
Cool Jupiter & $R_{\rm p} > 0.3$ $R_{\rm J}$ ~\&~ $P_{\rm{orb}} > 10$d & 48 & -1.94 $\pm$ 1.16 & 0.086\\
One close planet & \text{Multiplicity}=1 ~\&~ $P_{\rm{orb}} < 10$d & 197 & -1.54 $\pm$ 0.66 & 0.040\\
One distant planet & \text{Multiplicity}=1 ~\&~ $P_{\rm{orb}} > 10$d & 137 & -2.12 $\pm$ 0.78 & 0.012\\
\hline
\end{tabular}\\
{\raggedright The first two columns list the name of the sub-populations and their selection criteria (see text for details). Here, $P_{\rm{orb}}$ is the planetary orbital period and $R_p$ is the planetary radius ($R_J$ is the radius of Jupiter). Columns 3 and 4 list the number of stars of each sub-population and their average $\Delta P_{\rm{rot}}$, respectively. The final columns show the $p$-value of $\Delta P_{\rm{rot}}$ for a null hypothesis of $\Delta P_{\rm{rot}}=0$. \par}
\end{table*}

\begin{figure*}
    \centering
    \includegraphics[scale=.78]{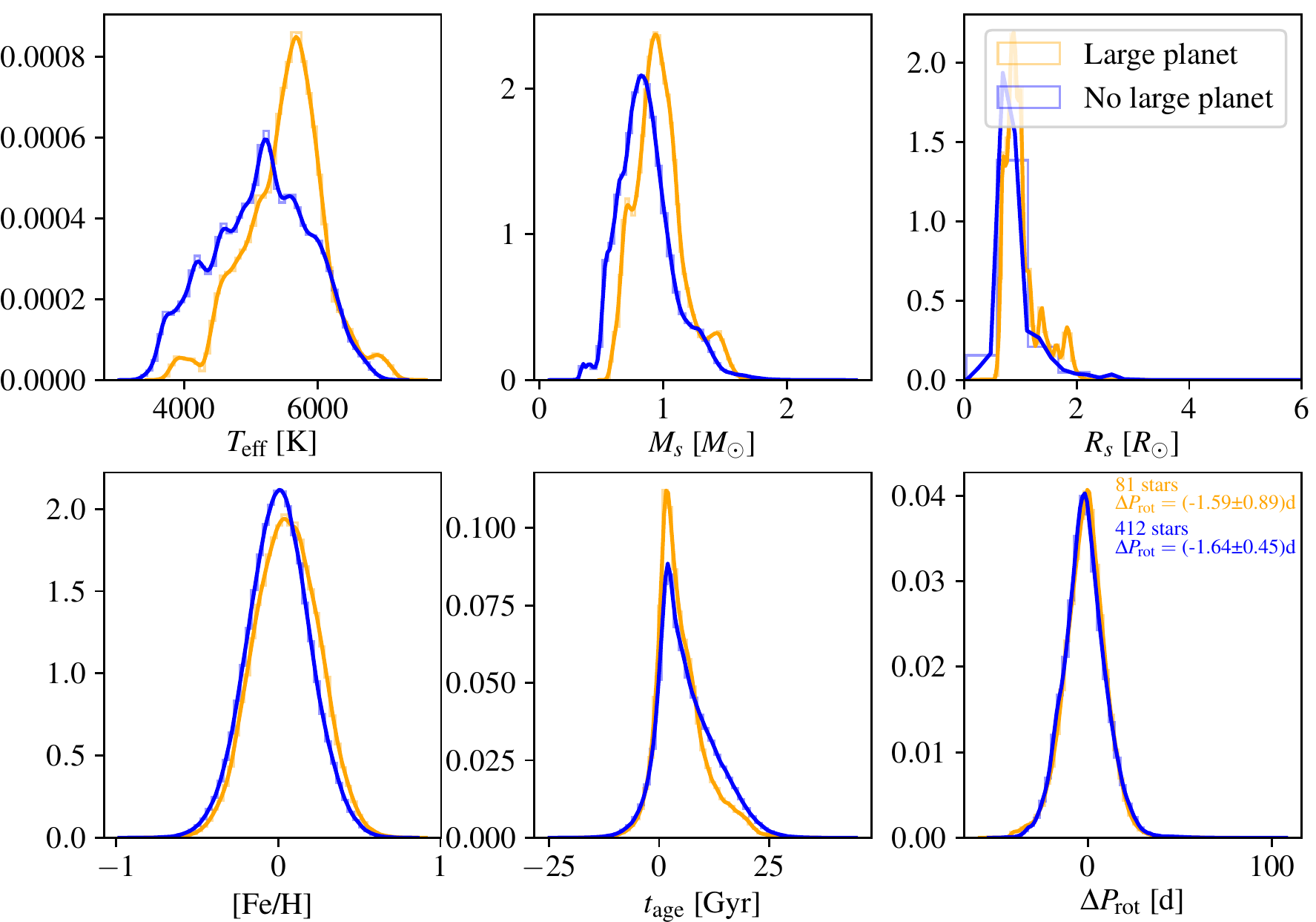}
    \caption{Stellar parameter histograms for the `large planet' (orange) and `no large planet' (blue) subpopulations. The amount of stars in and the $\Delta P_{\rm{rot}}$ of each one are indicated in the bottom-right panel. 
    }
    \label{fig:big vs small planet}
\end{figure*}
\begin{figure*}
    \centering
    \includegraphics[scale=.78]{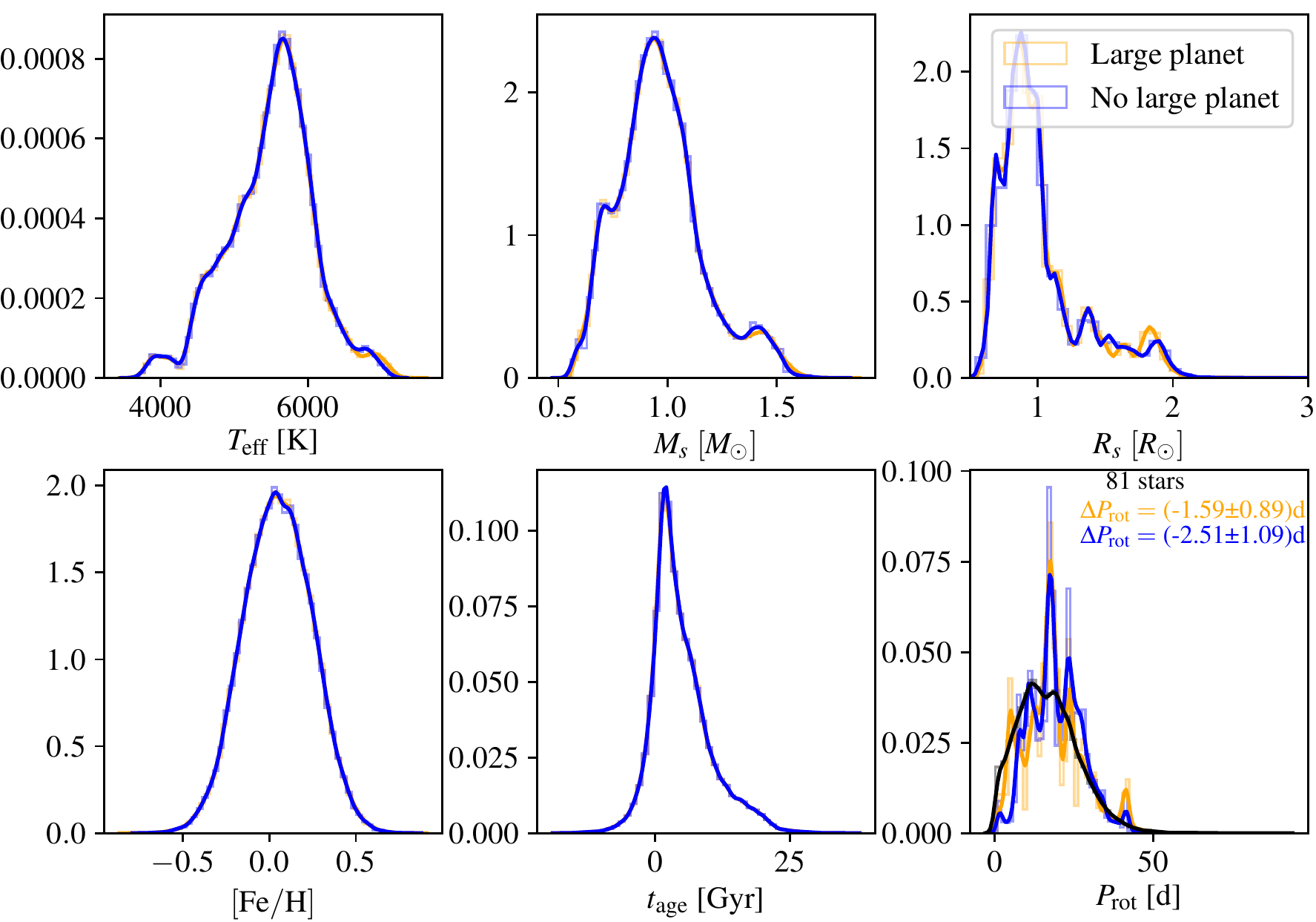}
    \caption{Stellar parameter histograms for the `large planet' (orange) and matched `no large planet' (blue) sub-populations. We indicate the amount of stars considered in the lower right panel ($P_{\rm{rot}}$). Also shown are the rotation periods of stars without planets matched to the same non-rotational stellar properties (black).}
    \label{fig:big vs small matched}
\end{figure*}

\begin{table}
\caption{Matching of the subpopulations.}
\label{table:planetary_param_matched}
\begin{tabular}{lll}
\hline
Subpopulation A & Subpopulation B & $\Delta P_{\rm{rot}}(B|A)$ [d]\\
\hline
Large planet & No large planet & $-2.51\pm1.09$\\
Multiple planets & One planet & $-1.55\pm0.86$\\
Distant planet & Close planet & $-1.97\pm0.63$\\
Hot Jupiter & Cool Jupiter & $-2.61\pm1.61$\\
One distant planet & One close planet & $-1.94\pm0.95$\\
\hline
\end{tabular}\\
{\raggedright Paired sub-populations of the dataset. In each line, sub-population A refers to the one containing the fewer stars and sub-population B to the one with the more.The $\Delta P_{\rm{rot}}$ of sub-population A and B can be found in \autoref{table:planetary_param}; we give here the $\Delta P_{\rm{rot}}$ of the sub-part of sub-population B that matches sub-population A ($\Delta P_{\rm{rot}}(B|A)$). \par}
\end{table}

Next, we compare the stellar properties of the various sub-samples. \autoref{fig:big vs small planet} shows the stellar properties of the `large planet' and `no large planet' sub-populations consisting of 81 and 412 stellar systems, respectively. On average, stellar hosts of large planets are hotter and more massive. 
The figure also highlights differences in the stellar age distribution. However, the distribution of $\Delta P_{\rm rot}$ 
(the offset between the rotational period of stars in the sub-population relative to a matched control sample) is similar between the two sub-samples.

In addition, we confirm that large planets are found around stars with higher metallicity than the stars hosting no large planets. This is in agreement with previous results that the occurrence rate of giant planets is enhanced around metal-rich stars \citep[see][among others]{Fischer2005,Santos2004,Santos2005,Alves2010}. Similar conclusions can be drawn when comparing stars with Hot or Cool Jupiters against other stars.

We provide additional figures to \autoref{fig:big vs small planet} in the appendix comparing the other sub-populations. Overall, we find some smaller difference for stars with multiple- vs single- planetary systems, distant vs no-distant planets, and single planetary systems with and without distant planets. However, we find significant differences between stars hosting Hot vs Cool Jupiters. In particular, stars with Hot Jupiters appear to be hotter and more massive, strongly peaking around Sun-like $T_{\rm eff}$ and $M_{\rm s}$, while Cool Jupiters have a much larger range in effective temperature and stellar mass.

However, comparing the difference in rotation periods of these sub-samples could be misleading if taken at face value due to correlations between non-rotational stellar properties and $P_{\rm rot}$ (see \autoref{fig:nonrot_vs_Prot} and \autoref{table:corr_coeffs}). Differences in non-rotational stellar properties of the sub-samples could have a physical origin or be related to selection effects and biases. For example, small planets are more difficult to detect around massive (and larger) stars which could result in a different distribution of stellar masses for samples with small vs large planets. Indeed, we find that stars hosting large planets tend to be more massive (and hotter) in comparison to stars hosting no large planets. Alternatively, this trend could reflect the underlying physics of the planetary formation process where giant planets are more likely to be formed around more massive stars.

Given differences in the distributions in non-rotational stellar properties in the sub-samples, correlations between non-rotational stellar properties and rotation periods need to be considered. For instance, as listed in \autoref{table:corr_coeffs}, there is a clear anti-correlation between the stellar rotation period and mass, temperature, and radius \citep[][]{Kraft1967}. In addition, we find a positive correlation between stellar rotation period and age \citep[][]{Skumanich1972}. Clearly, these correlations together with the differences in the distributions of non-rotational stellar properties in the sub-samples may affect the distribution of the inferred rotation periods.

To mitigate the role of these potential biases, we now compare our sub-samples when matching their non-rotational stellar properties as described in \autoref{subsec:matching}. 
\autoref{fig:big vs small matched} shows the distribution of the stellar properties for stars with large planets as well as for stars without large planets adjusted to have the same non-rotational stellar properties. Interestingly, compared with \autoref{fig:big vs small planet}, we find a noticeable difference in the $\Delta{}P_{\rm rot}$ of the two sub-samples. Specifically, stars with a large planet have $\Delta P_{\rm{rot}} = -1.59 \pm 0.89$d, while stars without a large planet have (after matching their non-rotational properties) $\Delta P_{\rm{rot}} = -2.51 \pm 1.1$d.
This change can be explained by the decrease in rotation period with increased stellar mass and effective temperature. Stars in our 'no large planet' sub-sample have on average smaller stellar masses than stars in the 'large planet' sub-sample. By matching, the average stellar mass in the 'no large planet' sub-sample is increased, which translates into a reduction in the stellar rotation period.
\autoref{fig:big vs small matched} also shows the distribution of $P_{\rm rot}$ of stars in the control sample with matched non-rotational stellar properties. As expected, stars without planets rotate on average slightly faster than either stars with or without a large planet.

\begin{figure}
    \centering
    \includegraphics[scale=.5]{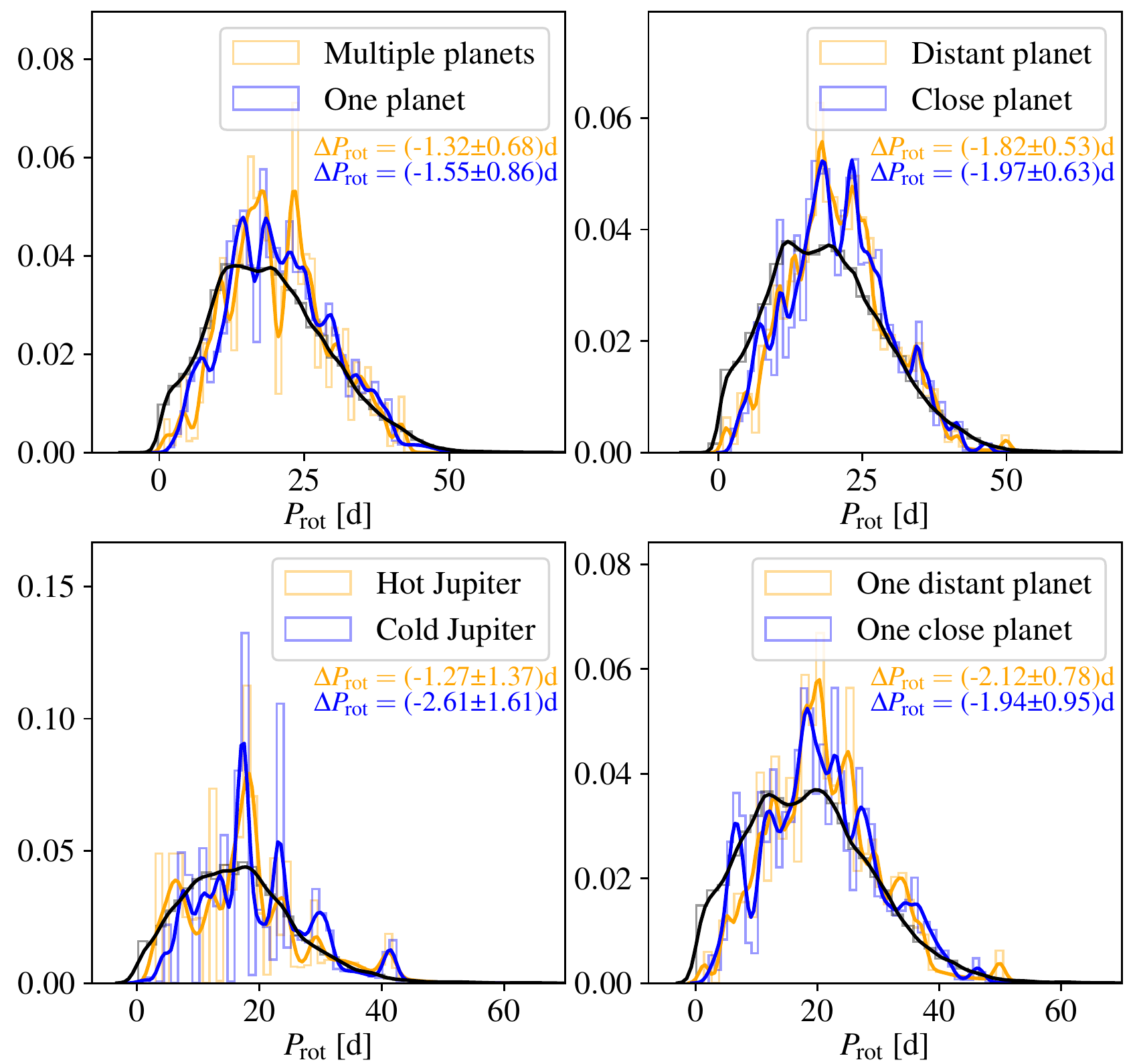}
    \caption{Distributions of rotation periods for matched subpopulations of \autoref{table:planetary_param_matched}.}
    \label{fig:matched_Prot_histograms}
\end{figure}

\autoref{fig:matched_Prot_histograms} shows the $P_{\rm rot}$ distributions for the other sub-populations, i.e., distant vs no-distant, single vs multiple, Hot Jupiter vs Cool Jupiter, one close vs one distant planet.
The different sub-populations show very similar distributions without major discernible differences. However, in all cases $P_{\rm rot}$ is shifted in comparison to stars without planets with matched non-rotational stellar properties.    

\section{Discussion \& Conclusions}
\label{sec:discussion}

In this work we have studied whether the presence of a detected planet is correlated with the rotation period of the host star. To this end, we cross-matched non-rotational stellar parameters derived by \citet{Berger2020} with stellar rotation periods measured in \citet{McQuillan2013b,McQuillan2014}. For stars with confirmed planets, we used planetary parameters from the \href{https://exoplanetarchive.ipac.caltech.edu/}{NASA exoplanet archive}. By carefully matching stars with and without detected planets on their non-rotational properties, we showed that planet-hosting stars rotate on average slower than stars without detected planets. We also investigated the dependence of this result on stellar and planetary properties. A possible application of this study is to more efficiently search for planet-hosts by targeting stars with longer rotation periods in future planet searches.

Our main results are summarised as follows:
\begin{itemize}
\item Stars with detected transiting planets rotate on average $1.63 \pm 0.40$ days slower than those without.

\item We confirm that stars hosting large planets have on average higher metallicities than stars hosting small  planets.

\item Sun-like stars show a significant difference in stellar rotation periods depending on the existence of a planet: $\Delta P_{\rm{rot}} = -2.53\pm0.50$ days for Sun-like masses ($\Delta P_{\rm{rot}} = -2.32\pm0.50$ days for Sun-like effective temperatures).

 
\item Stars with [Fe/H]$>0.1$ exhibit a significant $\Delta P_{\rm{rot}}$: $\Delta P_{\rm{rot}}=-2.66\pm0.85$ days.
\end{itemize}

This study, which represents only the first step in establishing a better understanding of the rotation periods of planet-hosting stars, comes with a few limitations.  
First, the control dataset of stars without planets may in fact  include stars that host undetected planets.
However, the potential existence of such false negatives in the control sample would only improve the statistical significance of our results. 
False negatives are planet-hosting stars in the control sample, and therefore their rotational periods are similar to the planet-hosting stellar sample. Consequently, the difference in rotation period between our planet-hosting stellar sample and the control sample would be even larger if false negatives were excluded. Some of the stars without planets may have had planets that were lost. This could have affected the rotational evolution of the stars. As a result, a better understanding of the physical processes taking place in young planetary systems and their effect on the stellar properties, as well as the efficiency of planet formation around different stellar types is desirable.  

Second, this study used catalogs of planets detected via the transit method by \emph{Kepler} \citep{Borucki2010} and in addition, currently we cannot exclude that the {\it Kepler} sample with detected planets has also some biases linked to the stellar rotation period. 
For example,  stellar activity as a proxy for rotation  may hinder the detection of planets by transits \citep[e.g., ][]{Bruno2021}. In this case we would expect that stars hosting smaller planets are more strongly affected by this selection bias, thus resulting in  lower average rotation periods compared to stars hosting larger planets. According to \autoref{table:planetary_param}, the rotation periods of such stellar hosts are comparable suggesting that such a selection bias is not dominant.
Future investigations based on catalogs of stellar hosts of planets detected by other methods, such as radial velocity, direct imaging, microlensing, and astrometry \citep[see ][for review]{Wright2013} which are subject to different observational biases, combined with the the statistical power of data collected by the {\it Gaia} mission \citep{GAIA2016} would allow to probe different parts of the stellar and planetary parameter spaces and to further test our findings.  

Third, determining stellar ages is challenging and this is reflected in the large measurement uncertainties of the ages in our samples. 
Stellar ages are important because on average older stars rotate more slowly \citep[e.g.,][]{Skumanich1972}. 
The uncertainty in age depends on the observational measurement approach \citep[see][for review]{Soderblom2010}. This paper uses age estimates based on isochrone fitting \citep{Berger2020} which has a relatively large uncertainty. 
A more accurate method for stellar age estimate using asteroseismology would decrease these uncertainties, as well as a careful consideration of the  bias in stellar rotation period and the detectability of asteroseismic oscillations. Indeed, it is more difficult to detect asteroseismic oscillations in magnetically active stars \citep{Chaplin2011,Mathur2019}, as a result asteroseismic samples are biased towards less active and thus slower-rotating stars.
In the future, the {\it PLATO} mission \citep{PLATO2014,PLATO2018} is expected to provide more accurate stellar ages using asteroseismology which would allow a more complete investigation accounting for the stellar ages. 

Fourth, we use a single (average) value for the rotation period for each star, ignoring potential differential rotation with radius and latitude. We suggest that future studies include differential rotation if rotation profiles are measured for a large number of stars.

While we focused on the stellar rotation in particular, our study more generally demonstrates the importance of investigating the relation between stellar properties and the existence of planets. 
Further research and more data are clearly desirable. In particular, it would be interesting to investigate whether other surveys using different detection methods (e.g., radial velocity) confirms the inferred trend.  
By having access to larger samples thanks to upcoming data, future studies will be able to better characterise the rotational and stellar properties of planet-hosting stars. 
This research direction is expected to blossom in the near future due to the large number of ongoing and future observations. For example, {\it Gaia} and {\it PLATO} will discover many new planets revealing new information on planets at larger radial distances and for hosts with various stellar properties.  
In addition, accurate measurements of stellar parameters such as rotation periods, ages, and metallicities will enhance our knowledge of the physical properties of planet-hosting stars and of the angular momentum distribution in stellar systems. This can further constrain planet formation models and improve our understanding of extra-solar planetary systems.

\section*{Acknowledgments}
The authors thank an anonymous referee for valuable comments. We also thank S. Udry and O. Attia for valuable discussions. RH acknowledges support from SNSF grant 200021\_169054. RF acknowledges financial support from the Swiss National Science Foundation (grant nos. 157591 and 194814). 

\section*{Data availability}
The data underlying this article will be shared on reasonable request. The datasets were derived from sources in the public domain: see \citet{Berger2020} for the non-rotational stellar properties, \citet{McQuillan2014,McQuillan2013b} for the stellar rotation periods, and the \href{https://exoplanetarchive.ipac.caltech.edu/}{NASA exoplanet archive} for the planets' properties.

\bibliographystyle{mnras}
\bibliography{References}


\appendix

\setcounter{figure}{0} \renewcommand{\thefigure}{A.\arabic{figure}}
\renewcommand{\theHfigure}{A.\thefigure}
\setcounter{table}{0} \renewcommand{\thetable}{A.\arabic{table}}
\setcounter{algorithm}{0} \renewcommand{\thealgorithm}{A.\arabic{algorithm}}

\section{Terminology} \label{Appendix1}
\begin{figure*}
    \centering
    \includegraphics[scale=.7]{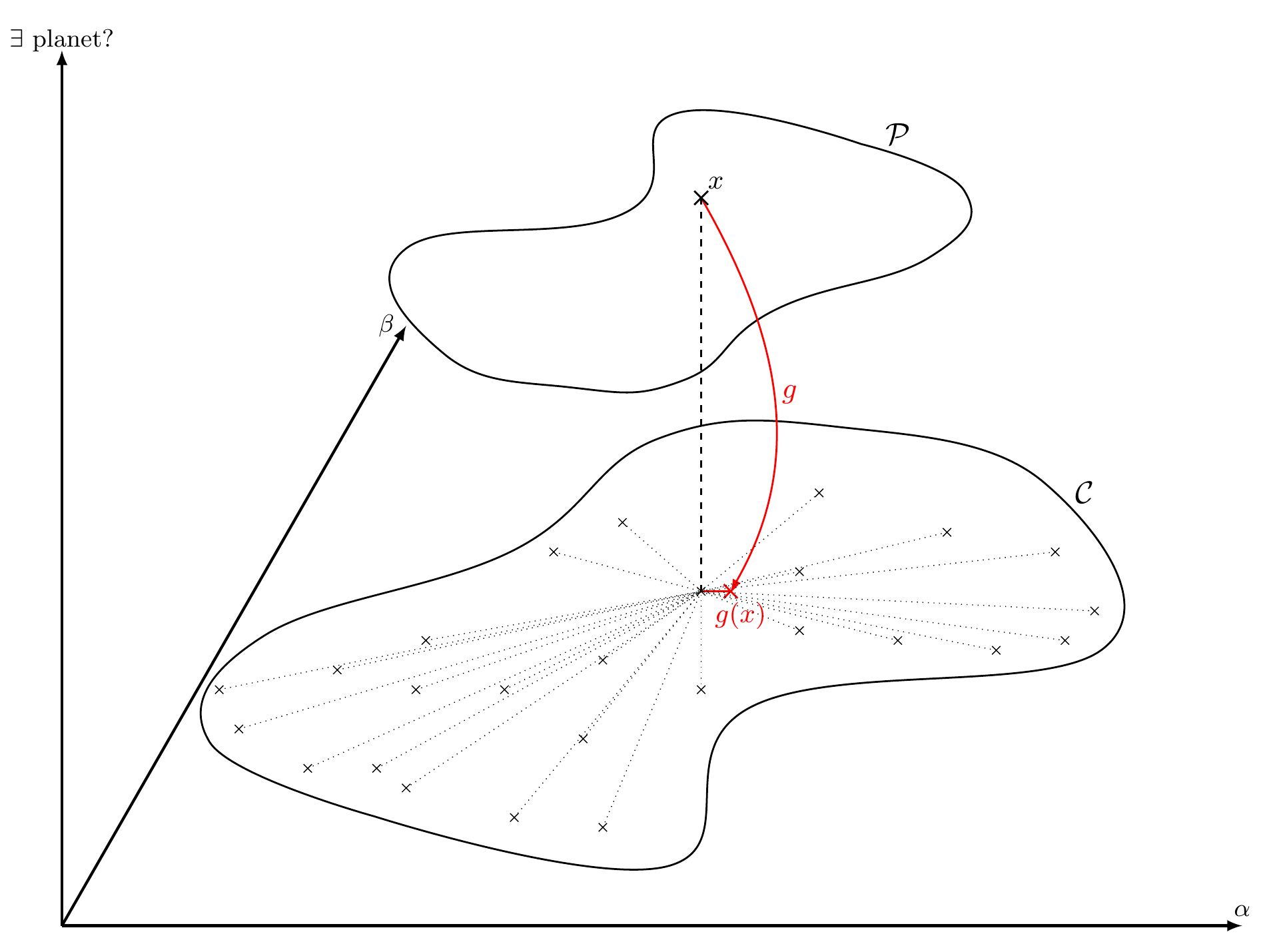}
    \caption{Diagram showing the matching process between the sets $\mathcal{P}^i$ and $\mathcal{C}^i$. We only show one matching operation; this is repeated for each $x^i \in \mathcal{P}^i$.}
    \label{fig:matching_diagram1}
\end{figure*}

\begin{figure*}
    \centering
    \includegraphics[scale=.7]{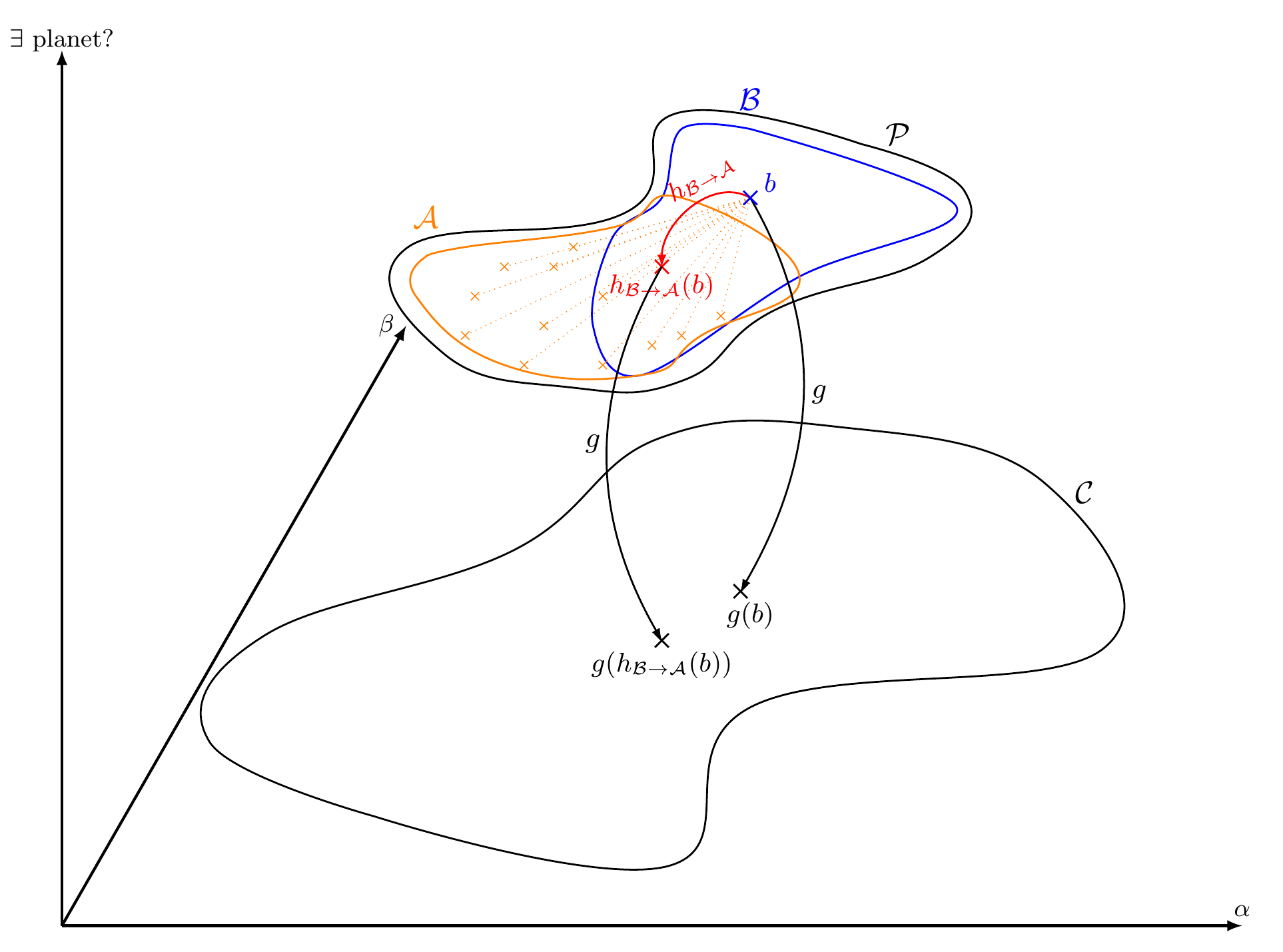}
    \caption{Diagram showing the matching process between two subsets $\mathcal{A}^i$ and $\mathcal{B}^i$ of $\mathcal{P}^i$. We only show one matching operation; this is repeated for each $b^i \in \mathcal{B}^i$.}
    \label{fig:matching_diagram2}
\end{figure*}

In this appendix we describe in detail the matching procedures described in the text.  
\autoref{fig:matching_diagram1} and \autoref{fig:matching_diagram2} illustrate the matching.\\
All the operations described here are repeated $N_{\rm{bootstrap}}$ (1000) times. We then compute the means and standard deviations, over all bootstraps, of the quantities we derive to obtain the values and errors we report in \autoref{sec:results}. The superscript $i$ indicates that the object is considered in the context of the i-th bootstrap iteration.

The datasets are defined as follows:
\begin{itemize}
    \item $\mathcal{P}$ is the set of stars with detected planets,
    \item $\mathcal{P}^i$ is the i-th bootstrap iteration of $\mathcal{P}$,
    \item $\mathcal{C}$ is the set of stars without detected planets.
    \item $\mathcal{C}^i$ is the i-th bootstrap iteration of $\mathcal{C}$,
\end{itemize}
One star $x^i$ is defined by its 6 stellar parameters, and $X^i$ is the column vector of its 5 non-rotational parameters:
\begin{equation}
    x=
    \begin{pmatrix}
    X=\begin{cases}
    T_{\rm{eff}} \\ M \\ R \\ \rm{[Fe/H]} \\ t_{\rm{age}} 
    \end{cases}
    \\ P_{\rm{rot}}
    \end{pmatrix}.
\end{equation}
We define a 5-D distance between two stars as the statistical distance between the non-rotational properties of the stars, weighted by the dispersion of those properties' values in the full $\mathcal{P}\cup\mathcal{C}$ dataset:
\begin{equation}
\begin{split}
    d_X(X_k,X_l)&=\left[
    \left(\frac{T_{\text{eff}_k}-T_{\text{eff}_l}}{\sigma_{T_{\rm{eff}}}}\right)^2
    +\left(\frac{M_k-M_l}{\sigma_{M}}\right)^2
    +\left(\frac{R_k-R_l}{\sigma_{R}}\right)^2
    \right. \\
    &+\left.
    \left(\frac{\text{[Fe/H]}_k-\text{[Fe/H]}_l}{\sigma_{\rm{[Fe/H]}}}\right)^2
    +\left(\frac{t_{\text{age}_k}-t_{\text{age}_l}}{\sigma_{t_{\rm{age}}}}\right)^2
    \right]^\frac{1}{2},
\end{split}
\label{eqn:app:d_X}
\end{equation}
where the subscripts $k$ and $l$ refer to the stars whose distance is being computed, and $\sigma_{T_{\rm{eff}}}$, $\sigma_{M}$, $\sigma_{R}$, $\sigma_{\text{[Fe/H]}}$, $\sigma_{t_{\rm{age}}}$ are the statistical standard deviations of each parameter over the big ($\mathcal{P} \cup \mathcal{C}$) dataset.\\

Let $g^i$ be the function that assigns, to each element of $\mathcal{P}^i$, its most similar counterpart in $\mathcal{C}^i$:
\begin{equation}
\begin{split}
g^i\colon & \mathcal{P}^i \to \mathcal{C}^i\\
          & x^i \mapsto \displaystyle\argmin_{c^i \in \mathcal{C}^i} d_X (x^i,c^i),
\end{split}
\end{equation}
where $d_X$ is the 5-D distance defined in \autoref{eqn:app:d_X}.
Then the multiset (because different stars in $\mathcal{P}^i$ can have the same counterpart, in which case this counterpart appears more than once in the multiset) $\mathcal{M}^i$, comprising of the matched counterparts in $\mathcal{C}^i$ to every star in $\mathcal{P}^i$, is:
\begin{equation}
    \mathcal{M}^i=\left\{g^i(x^i) | x^i \in \mathcal{P}^i \right\}.
\end{equation}

Let $\mathcal{A}^i$ and $\mathcal{B}^i$ be two subsets of $\mathcal{P}^i$. These correspond in practice to sub-populations based on specific stellar or planetary parameters.\\
Let $h_{\mathcal{B}^i\to\mathcal{A}^i}^i$ be the function that assigns, to each element of $\mathcal{B}^i$, its most similar counterpart in $\mathcal{A}^i$:
\begin{align}
\begin{split}
    h_{\mathcal{B}^i\to\mathcal{A}^i}^i\colon & \mathcal{B}^i \to \mathcal{A}^i\\
            & b^i \mapsto \displaystyle\argmin_{a^i \in \mathcal{A}^i} d_X (b^i,a^i).
\end{split}
\end{align}
Then the multiset (for reasons similar to $\mathcal{M}^i$) $\mathcal{A}^i | \mathcal{B}^i$ comprising of the matched counterparts in $\mathcal{A}^i$ to every star in $\mathcal{B}^i$ is:
\begin{equation}
    \mathcal{A}^i|\mathcal{B}^i=\left\{h_{\mathcal{B}^i\to\mathcal{A}^i}^i(b^i)|b^i\in \mathcal{B}^i \right\}.
\end{equation}

For any subset $\mathcal{S}^i$ of $\mathcal{P}^i$, let $\Delta P_{\rm{rot}}(\mathcal{S}^i)$ be the average period difference between the stars in $\mathcal{S}^i$ and their matched counterparts:
\begin{equation}
    \Delta P_{\rm{rot}} (\mathcal{S}^i)= \langle P_{\rm{rot}}(g^i(s^i)) - P_{\rm{rot}} (s^i)\rangle_{s^i\in \mathcal{S}^i}.
\end{equation}
Thus, the average period difference between the stars in $\mathcal{A}^i | \mathcal{B}^i$ is:
\begin{equation}
    \Delta P_{\rm{rot}} (\mathcal{A}^i|\mathcal{B}^i)= \langle P_{\rm{rot}} (g^i(h_{\mathcal{B}^i \to \mathcal{A}^i}^i (b^i))) - P_{\rm{rot}} (h_{\mathcal{B}^i \to \mathcal{A}^i}^i (b^i)) \rangle_{b^i \in \mathcal{B}^i}.
\end{equation}
Finally, let $\Delta\Delta P_{\rm{rot}}(\mathcal{A}^i | \mathcal{B}^i)$ be the difference in $\Delta P_{\rm{rot}}$ between $\mathcal{B}^i$ and $\mathcal{A}^i|\mathcal{B}^i$:
\begin{equation}
    \Delta\Delta P_{\rm{rot}}(\mathcal{A}^i | \mathcal{B}^i)=\Delta P_{\rm{rot}}(\mathcal{B}^i)-\Delta P_{\rm{rot}}(\mathcal{A}^i|\mathcal{B}^i).
\end{equation}
All the objects we defined here are confined to a certain bootstrap iteration. To get to our final $\Delta P_{\rm{rot}}$ and $\Delta\Delta P_{\rm{rot}}$ results, we compute the mean and standard deviation of those quantities over all bootstrap iterations: for any subset $\mathcal{S}$ of $\mathcal{P}$, whose $N_{\rm{bootstrap}}$ bootstrap iterations form the set $\left\{\mathcal{S}^i\right\}_{1\leq i \leq N_{\rm{bootstrap}}}$:
\begin{align}
\centering
    \overline{\Delta P_{\rm{rot}}(\mathcal{S})}&=\langle \Delta P_{\rm{rot}}(\mathcal{S}^i) \rangle_{1\leq i \leq N_{\rm{bootstrap}}}\\
    e(\Delta P_{\rm{rot}}(\mathcal{S}))&= \sigma(\Delta P_{\rm{rot}}(\mathcal{S}^i))_{1\leq i \leq N_{\rm{bootstrap}}}\\
    \Delta P_{\rm{rot}}(\mathcal{S})&=(\overline{\Delta P_{\rm{rot}}(\mathcal{S})}\pm e(\Delta P_{\rm{rot}}(\mathcal{S}))
\end{align}

\renewcommand{\algorithmiccomment}[1]{\hfill \small//~#1\normalsize}
\begin{algorithm}[H]
\caption{Bootstrapping and matching processes}
\label{algo1}
\begin{algorithmic}
\STATE \textbf{\emph{Bootstrapping}}
\FOR{bootstrap iteration i in $(1,1000)$}
    \FOR{$x\in\text{Full dataset}$}
        \STATE $x^i \leftarrow x + \sigma_x \mathcal{N}(0,1)$
    \ENDFOR
\ENDFOR
\vspace{5pt}
\STATE \textbf{\emph{Matching}}
\FOR{bootstrap iteration i in $N_{\rm{bootstrap}}(1,1000)$}
    \FOR{$x_k^i \in \text{Stars with planets}^i$}
        \FOR{$x_l^i \in \text{Stars without planets}^i$}
            \STATE $d_X (x^i_k,x^i_l) \leftarrow \sqrt{\left(\frac{X_k^i-X_l^i}{\sigma_X}\right)^2}$
        \ENDFOR
        \STATE $x_k^{\prime i}\leftarrow \displaystyle\argmin_{x_l^i ~\text{without planet}} d_X (x_k^i,x_l^i)$ \COMMENT{$x_k^{\prime i}$ is $x_k^i$'s closest counterpart in iteration i}
        \STATE $\Delta P_{\rm{rot}} (x_k^i) \leftarrow P_{\rm{rot}}(x_k^{\prime i}) - P_{\rm{rot}}(x_k^i)$
    \ENDFOR
    \STATE $\Delta P_{\rm{rot}}^i \leftarrow \left< \Delta P_{\rm{rot}} (x_k^i) \right>_{x_k^i ~\text{with planet}}$
\ENDFOR
\STATE $\Delta P_{\rm{rot}} \leftarrow \left(\left<\Delta P_{\rm{rot}}^i\right>_{1\leq i \leq 1000} \pm \sigma(\Delta P_{\rm{rot}}^i)_{1\leq i \leq 1000}\right)$
\end{algorithmic}
\end{algorithm}
\begin{table}
\caption{$\Delta P_{\rm rot}$ values for stars older and younger than certain ages. \hspace*{300pt}}
\label{table:deltas_over_under}
\begin{tabular}{lcccc}
\hline
Age & $N_{\rm stars}$ younger & $\Delta P_{\rm{rot}}$ [d] younger & $N_{\rm stars}$ older & $\Delta P_{\rm{rot}}$ [d] older\\
\hline
1Gyr & 12 & $-1.58\pm1.30$ & 481 & $-1.71\pm0.38$\\
1.5Gyr & 31 & $-2.53\pm0.92$ & 462 & $-1.58\pm0.41$\\
2Gyr & 64 & $-2.48\pm0.75$ & 429 & $-1.53\pm0.45$\\
2.5Gyr & 95 & $-1.77\pm0.66$ & 397 & $-1.60\pm0.47$\\
3Gyr & 133 & $-1.55\pm0.57$ & 359 & $-1.64\pm0.50$\\
\hline
\end{tabular}
\end{table}
\begin{figure}
    \centering
    \includegraphics[scale=.5]{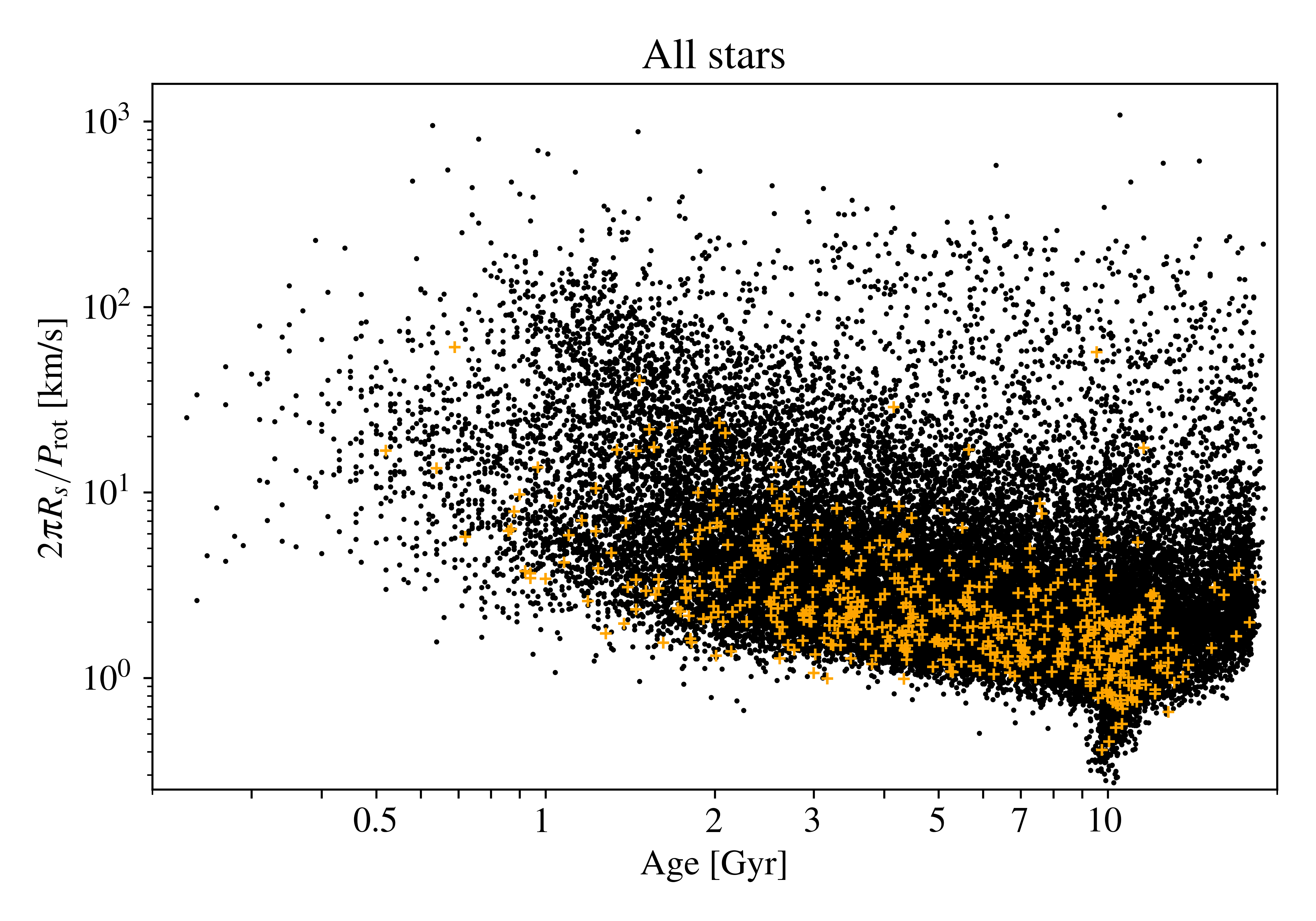}
    \caption{Scatter plot of all the stars in the two main datasets of our paper (control: black, stars with planets: orange) in the $t_{\rm age} - v_{\rm rot}$ plane. Note the logarithmic scale of the axes.}
    \label{fig:Age_vrot_All}
\end{figure}
\begin{figure*}
    \centering
    \includegraphics{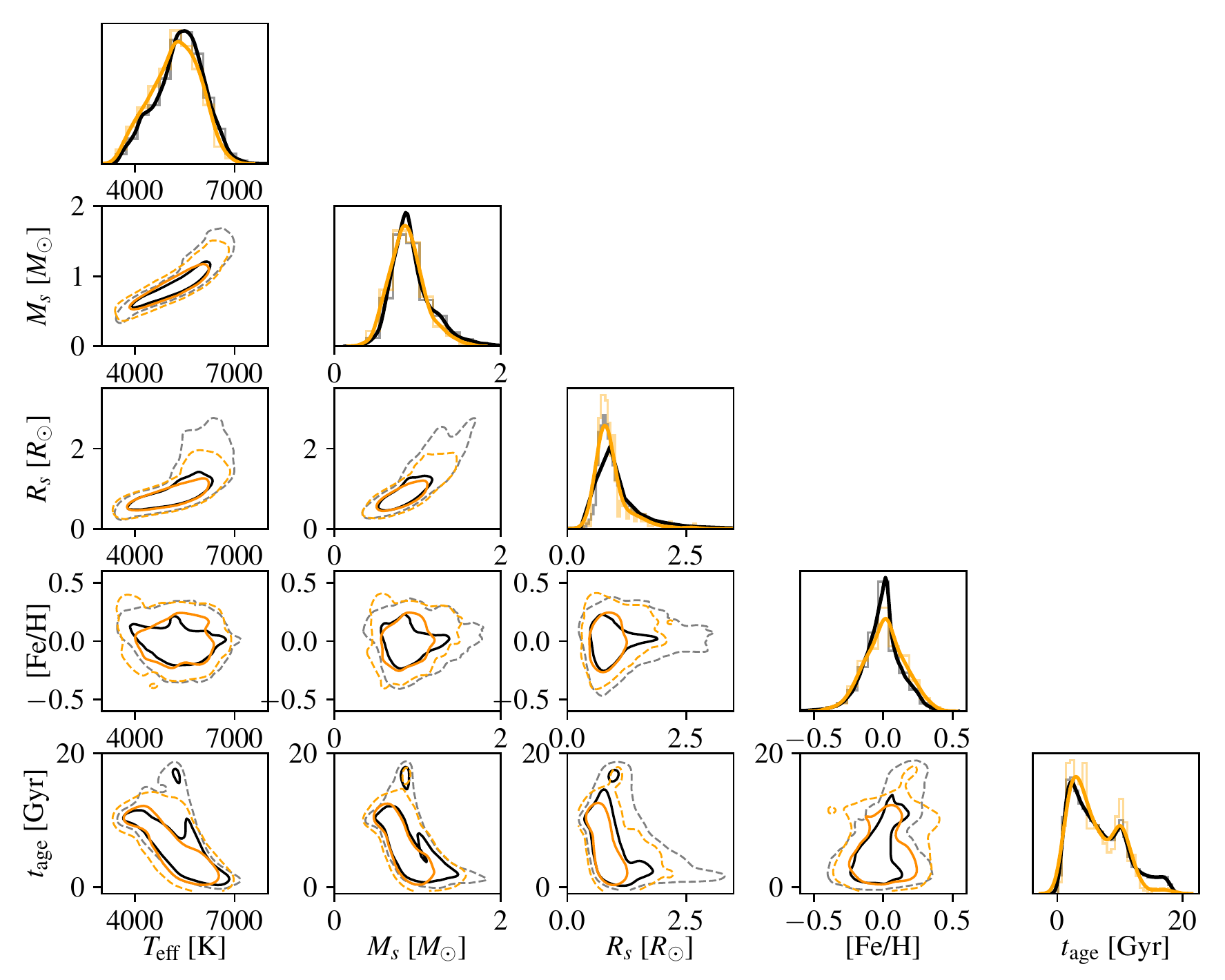}
    \caption{Same as \autoref{fig:pairplots}, but before matching. Diagonal plots show one-dimensional probability distribution functions of the two data sets, while off-diagonal plots show two-dimensional distributions. In each panel, orange (black) lines and points represent  stars with (without) detected planets. Contour lines encircle 68\% (solid) and 95\% (dashed) of all the stars.}
    \label{fig:pairplots_prematch}
\end{figure*}

\begin{table*}
\caption{All subpopulations including different parameter cutoff values. \hspace*{170pt}}
\label{table:appendix_table}
\begin{tabular}{lclll}
\hline
Name & Filter & $N_{\rm{stars}}$ & $\Delta P_{\rm{rot}}$ [d] & p-value\\
\hline
Far planet & $P_{\rm{orb}}$ > 10d & 265 & $-1.82 \pm 0.53$ & 0.0021\\
Far planet2 & $P_{\rm{orb}}$ > 25d\text{ (average period in the dataset)} & 122 & $-2.18 \pm 0.74$ & 0.0070\\
No far planet & $P_{\rm{orb}}$ < 10d & 228 & $-1.41 \pm 0.61$ & 0.045\\
No far planet2 & $P_{\rm{orb}}$ < 25d & 371 & $-1.45 \pm 0.47$ & 0.0073\\
Large planet & $R_p$ > 0.3 $R_J$ & 81 & $-1.59 \pm 0.89$ & 0.092\\
Large planet2 & $R_p$ > 0.187 $R_J$\text{ (median radius in the dataset)} & 276 & $-1.84 \pm 0.52$ & 0.0015\\
Large planet3 & $R_p$ > 0.233 $R_J$\text{ (average radius in the dataset)} & 175 & $-1.54 \pm 0.63$ & 0.032\\
No large planet & $R_p$ < 0.3$R_J$ & 412 & $-1.64 \pm 0.45$ & 0.0013\\
No large planet2 & $R_p$ < 0.187$R_J$\text{ (median radius in the dataset)} & 217 & $-1.37 \pm 0.64$ & 0.063\\
No large planet3 & $R_p$ < 0.233$R_J$\text{ (average radius in the dataset)} & 318 & $-1.68 \pm 0.53$ & 0.0050\\
Single planet & \text{Multiplicity}=1 & 334 & $-1.78 \pm 0.5$ & 0.0014\\
Multiple planets & \text{Multiplicity} > 1& 159 & $-1.32 \pm 0.68$ & 0.091\\
Hot Jupiter & $R_p$ > 0.3 $R_J ~\&~ P_{\rm{orb}}$ < 10d & 37 & $-1.27 \pm 1.37$ & 0.19\\
Hot Jupiter2 & $R_p$ > 0.187 $R_J ~\&~ P_{\rm{orb}}$ < 10d & 121 & $-1.92 \pm 0.77$ & 0.023\\
Hot Jupiter3 & $R_p$ > 0.233 $R_J ~\&~ P_{\rm{orb}}$ < 10d & 75 & $-1.58 \pm 0.95$ & 0.11\\
Hot Jupiter4 & $R_p$ > 0.36 $R_J ~\&~ P_{\rm{orb}}$ < 10d & 30 & $-1.66 \pm 1.52$ & 0.14\\
Cool Jupiter & $R_p$ > 0.3 $R_J ~\&~ P_{\rm{orb}}$ > 10d & 48 & $-1.94 \pm 1.16$ & 0.086\\
Cool Jupiter2 & $R_p$ > 0.187 $R_J ~\&~ P_{\rm{orb}}$ > 10d & 187 & $-1.80 \pm 0.64$ & 0.012\\
Cool Jupiter3 & $R_p$ > 0.233 $R_J ~\&~ P_{\rm{orb}}$ > 10d & 116 & $-1.62 \pm 0.76$ & 0.054\\
Cool Jupiter4 & $R_p$ > 0.36 $R_J ~\&~ P_{\rm{orb}}$ > 10d & 32 & $-2.07 \pm 1.45$ & 0.099\\
One close planet & $\text{Multiplicity}=1 ~\&~ P_{\rm{orb}}$ < 10d & 197 & $-1.54 \pm 0.66$ & 0.040\\
One distant planet & $\text{Multiplicity}=1 ~\&~ P_{\rm{orb}}$ > 10d & 137 & $-2.12 \pm 0.78$ & 0.012\\
Low-mass&$M_s < 0.8M_\odot$ & 183 & $-0.35 \pm 0.75$ & 0.48\\
Sun-like mass&$0.8M_\odot < M_s < 1.2M_\odot$ & 267 & $-2.53 \pm 0.50$ & $2.2\times10^{-6}$\\
High-mass&$M_s > 1.2M_\odot$ & 43 & $-1.52 \pm 1.12$ & 0.075\\
Cool&$T_{\rm{eff}}$ < 5000K & 181 & $-0.63 \pm 0.78$ & 0.37\\
Sun-like effective temperature&5000K < $T_{\rm{eff}}$ < 6000K & 244 & $-2.34 \pm 0.50$ & $1.4\times10^{-5}$\\
Hot&$T_{\rm{eff}}$ > 6000K & 68 & $-1.77 \pm 0.86$ & 0.056\\
Low metallicity&\text{[Fe/H]} < -0.1 & 106 & $-1.43 \pm 0.80$ & 0.10\\
Solar metallicity&-0.1 < \text{[Fe/H]} < 0.1 & 257 & $-1.19 \pm 0.54$ & 0.065\\
High metallicity&\text{[Fe/H]} > 0.1 & 130 & $-2.66 \pm 0.85$ & 3.5 $\times 10^{-3}$\\
Young stars&$t_{\rm{age}} < 5\text{Gyr}$ & 244 & $-1.77 \pm 0.48$ & $9.3\times10^{-4}$\\
Intermediate age&$5\text{Gyr} < t_{\rm{age}} < 8\text{Gyr}$ & 98 & $-1.89 \pm 0.93$ & 0.054\\
Old stars&$t_{\rm{age}} > 8\text{Gyr}$ & 151 & $-1.24 \pm 0.86$ & 0.17\\
\end{tabular}\\
{\raggedright Sub-populations of planet-hosting stars based on planetary properties (similar to \autoref{table:planetary_param} but including all the parameter cutoff values we used to make sub-populations). \par}
\end{table*}

\begin{figure*}
    \centering
    \includegraphics[scale=.8]{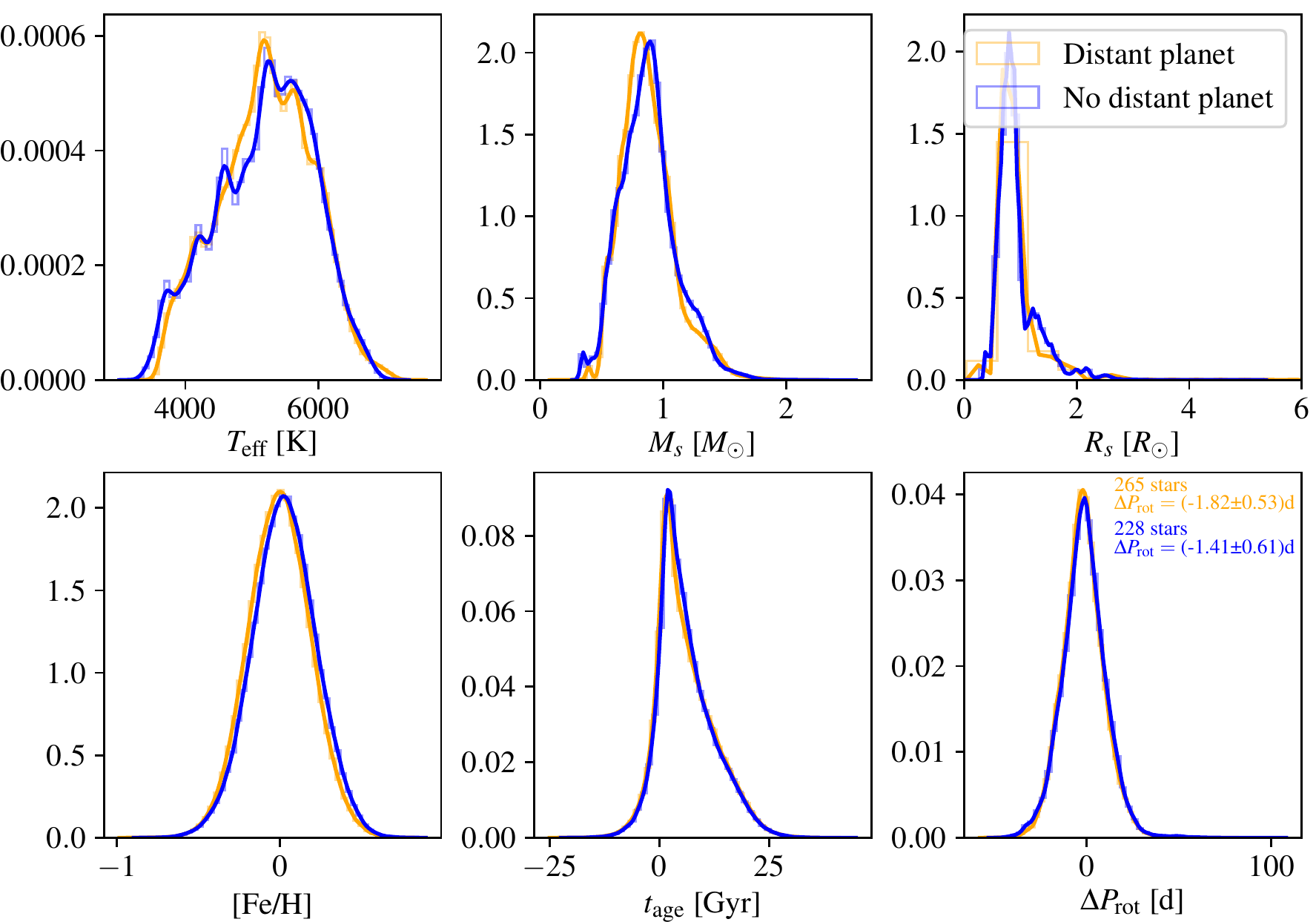}
    \caption{Stellar parameter histograms for the `distant planet' (orange) and `no distant planet' (blue) subpopulations. The amount of stars in and the $\Delta P_{\rm{rot}}$ of each one are indicated in the bottom-right panel.}
    \label{fig:distant_vs_close}
\end{figure*}

\begin{figure*}
    \centering
    \includegraphics[scale=.8]{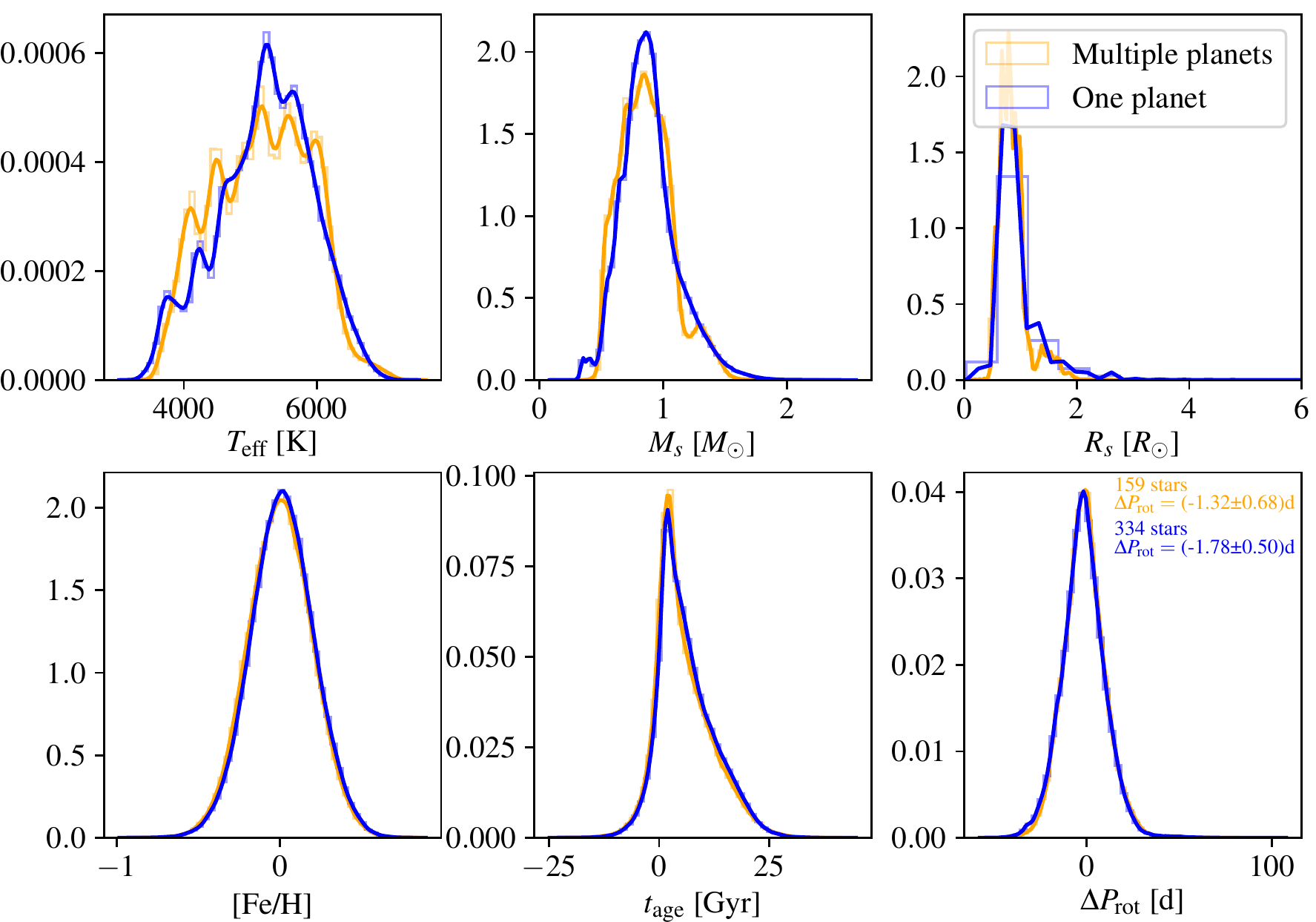}
    \caption{Stellar parameter histograms for the `multiple planets' (orange) and `single planet' (blue) subpopulations. The amount of stars in and the $\Delta P_{\rm{rot}}$ of each one are indicated in the bottom-right panel.}
    \label{fig:multiple_vs_one}
\end{figure*}

\begin{figure*}
    \centering
    \includegraphics[scale=.8]{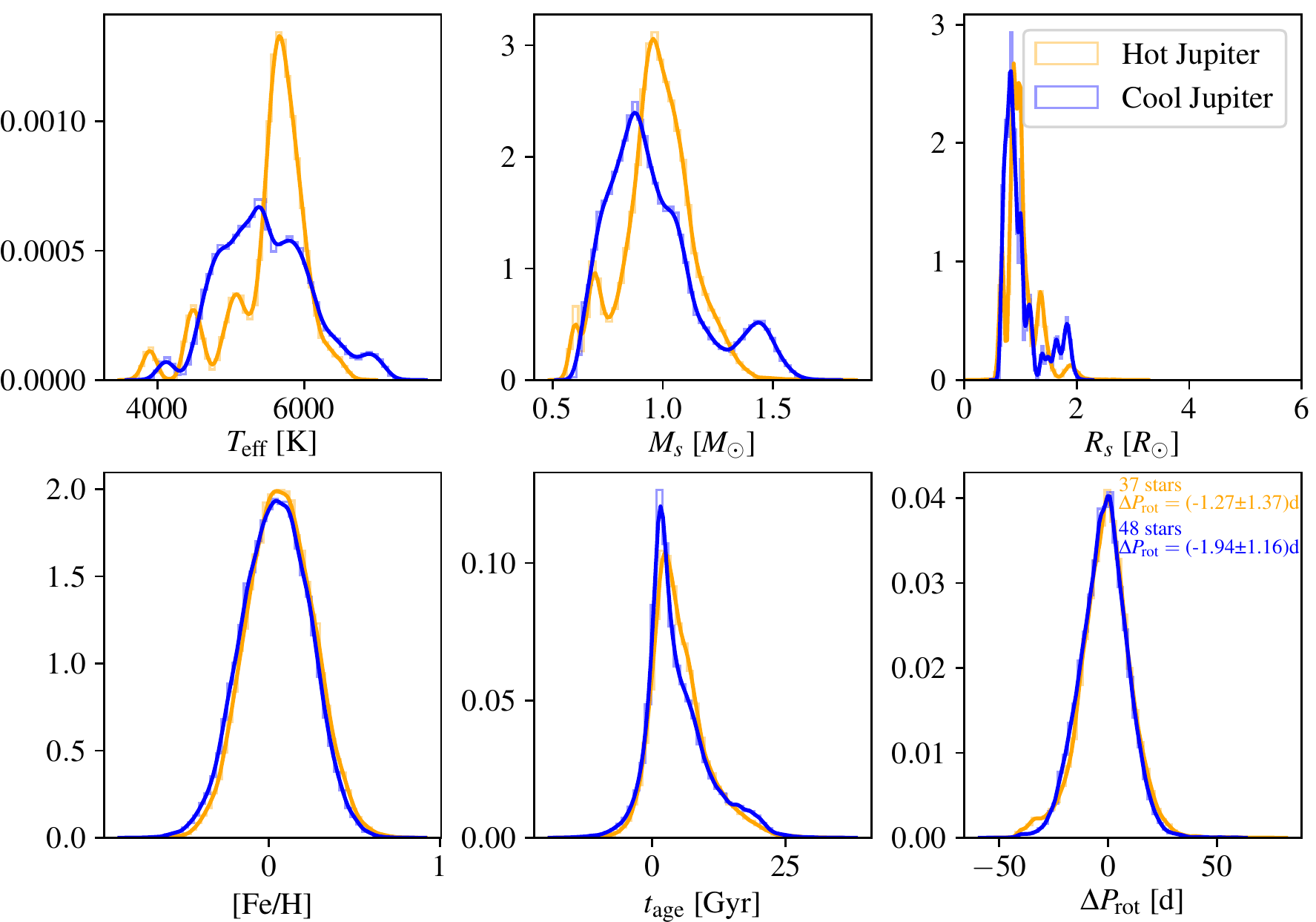}
    \caption{Stellar parameter histograms for the `Hot Jupiter' (orange) and `Cool Jupiter' (blue) subpopulations. The amount of stars in and the $\Delta P_{\rm{rot}}$ of each one are indicated in the bottom-right panel.}
    \label{fig:HJ_vs_CJ}
    \centering
    \includegraphics[scale=.8]{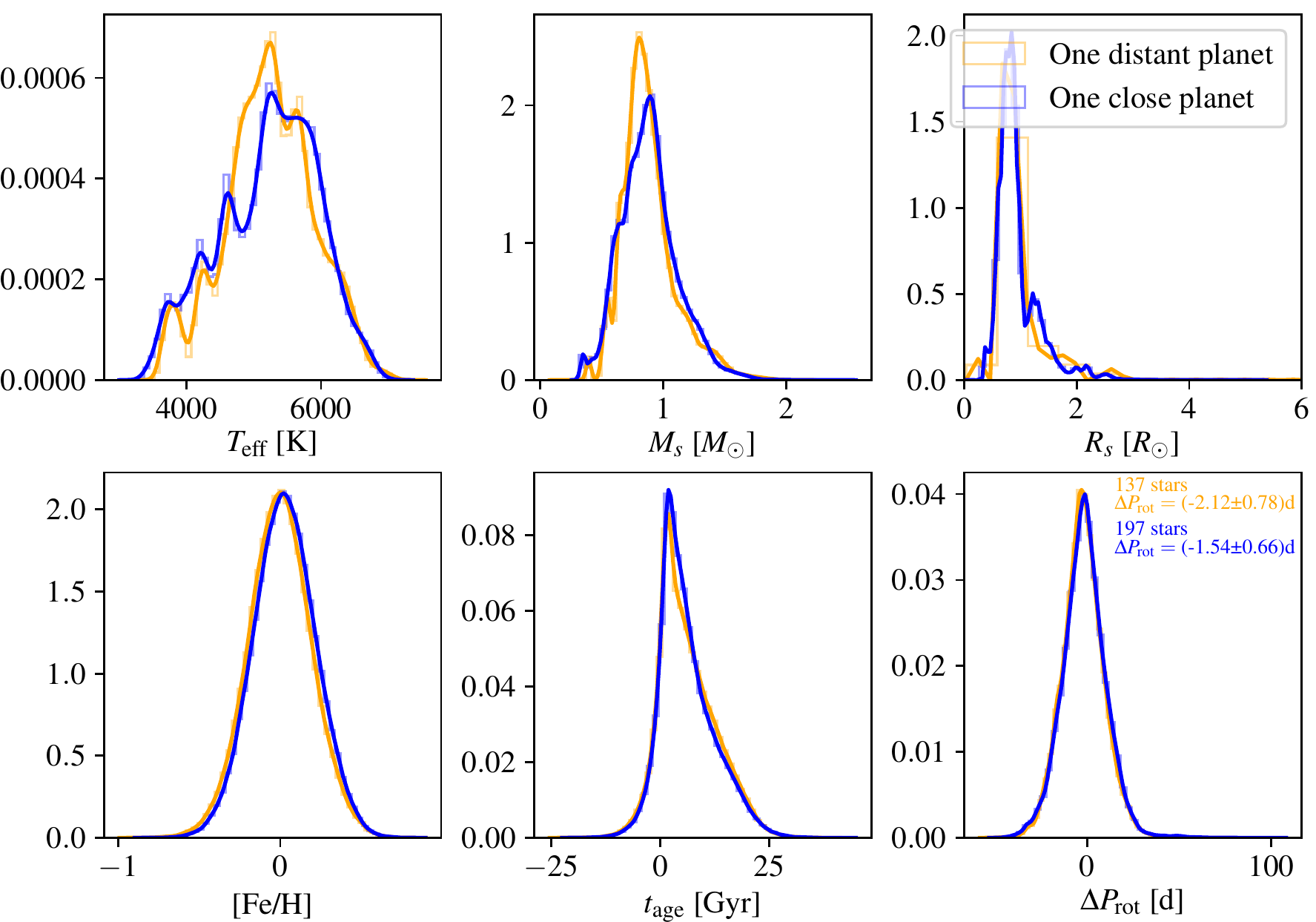}
    \caption{Stellar parameter histograms for the `one distant planet' (orange) and `one close planet' (blue) subpopulations. The amount of stars in and the $\Delta P_{\rm{rot}}$ of each one are indicated in the bottom-right panel.}
    \label{fig:oneclose_vs_onedistant}
\end{figure*}

\clearpage

\setcounter{figure}{0} \renewcommand{\thefigure}{B.\arabic{figure}}
\setcounter{table}{0} \renewcommand{\thetable}{B.\arabic{table}}
\setcounter{algorithm}{0} \renewcommand{\thealgorithm}{B.\arabic{algorithm}}
\setcounter{equation}{0} \renewcommand{\theequation}{B.\arabic{equation}}
\renewcommand{\theHfigure}{B.\thefigure}

\section{Validation of the matching approach} \label{Appendix2}

In order to validate our method, we apply it to a biased sample of stars without detected planets. For the purpose of this test, this sample is randomly drawn from a parent sample with selection probabilities that depend on non-rotational stellar properties. The so-constructed sample is generally biased with respect to both non-rotational and rotational stellar properties. The (unbiased) parent sample serves as the control sample. 
\par
The goal of our matching approach is to remove relative biases between the two samples to identify actual differences in rotation. Since, in this test, the biased sample is selected from the parent sample without considering the stellar rotation period, we expect to find no significant difference in rotation after matching.

We test our method using a synthetically created biased sample since no adequate samples are available. For instance, asteroseismic surveys, which provide accurate stellar rotation periods of stars, are strongly biased towards less magnetically active stars, and as a result slowly rotating stars. Hence, even after matching, the rotation periods in such a sample would generally be longer than the rotation periods of a representative sample of stars without planets.

\begin{figure*}
    \centering
    \includegraphics[scale=.8]{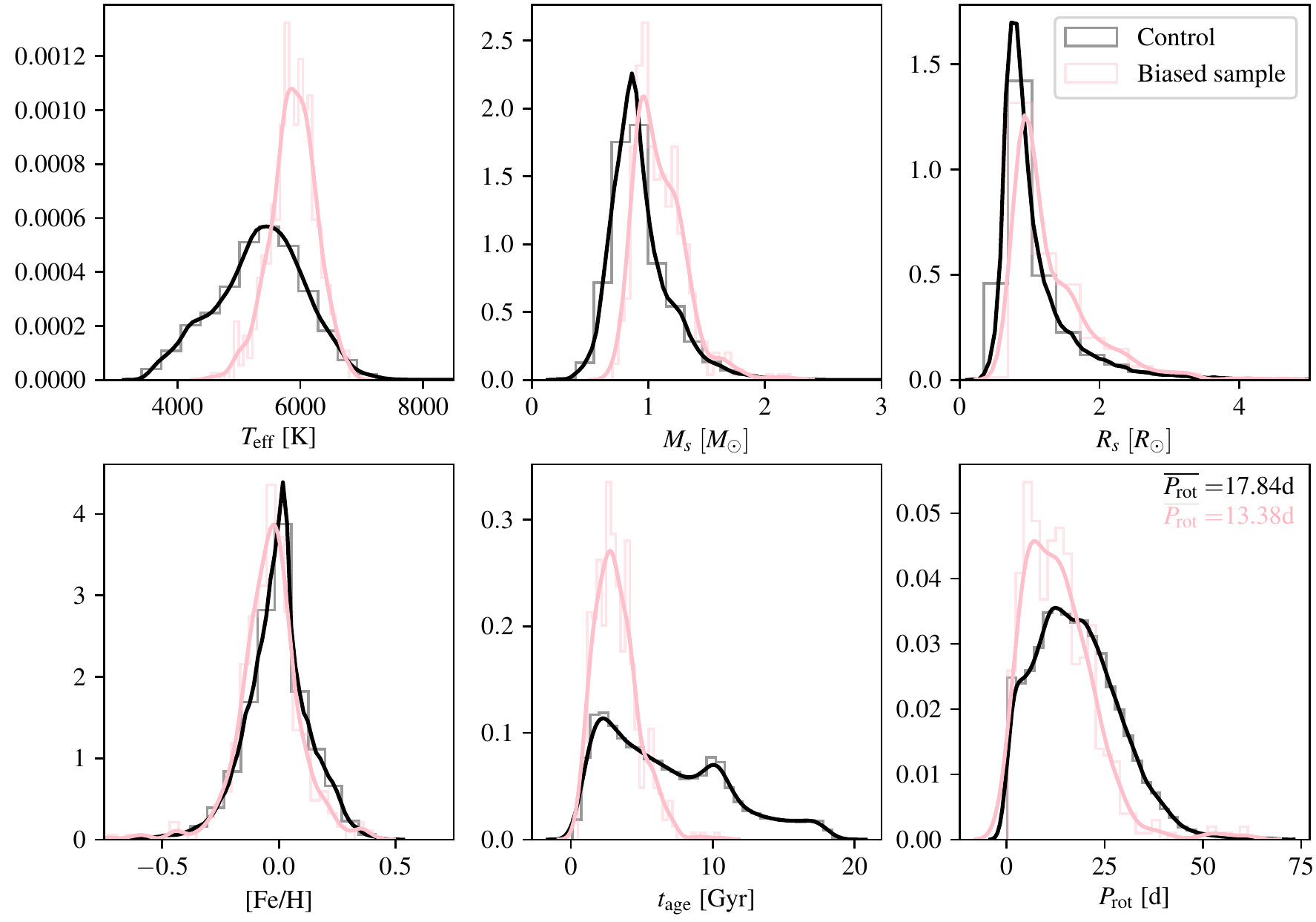}
    \caption{Histograms of the stellar properties of our control sample (shown in black) and a biased random sample (shown in pink).
    \label{fig:randsamp_prematch}}
    \includegraphics[scale=.8]{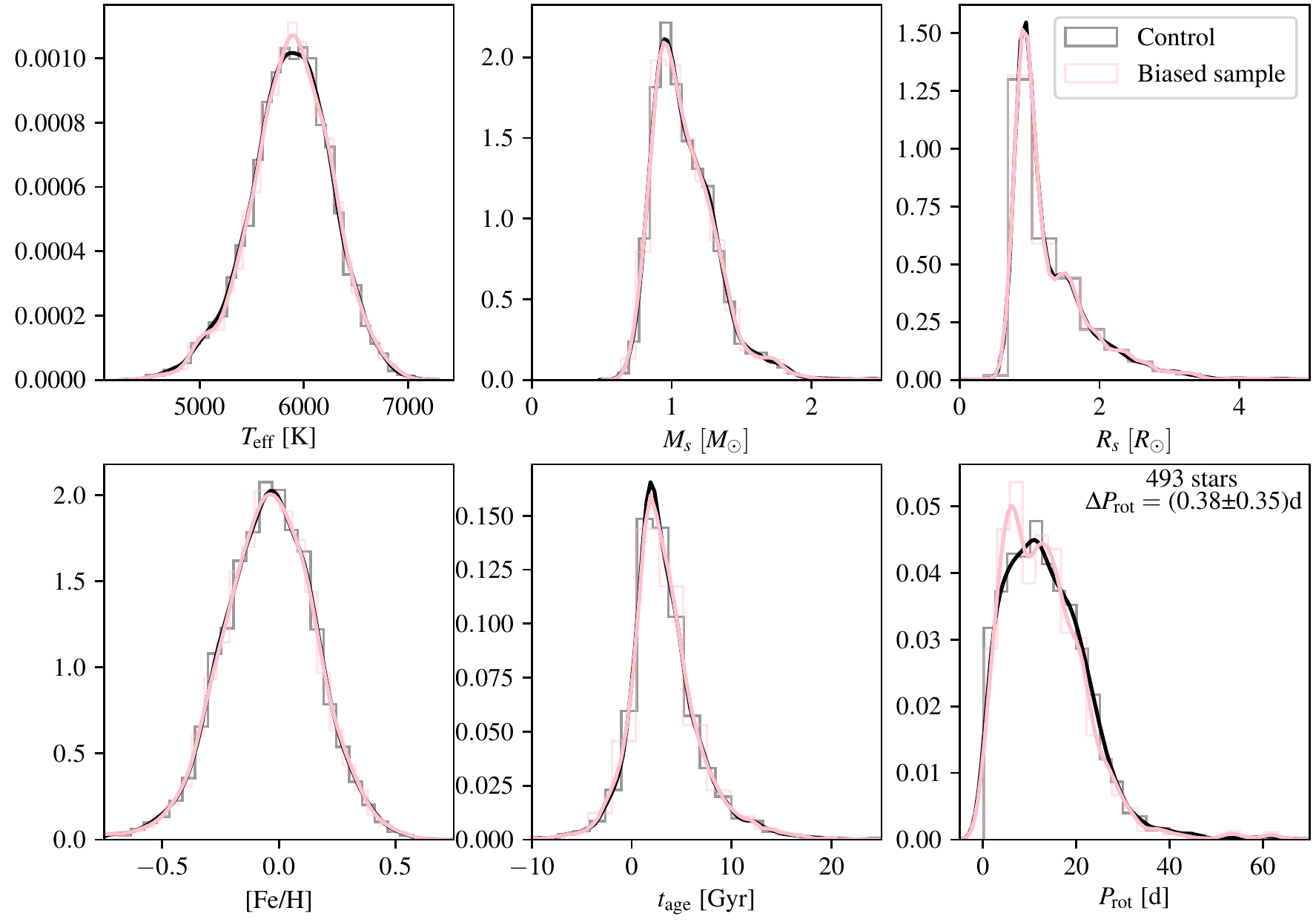}
    \caption{Histograms of the stellar properties after matching the control (shown in black) to the biased random sample (shown in pink). 
    \label{fig:randsamp_postmatch}}
\end{figure*}

We create the biased sample by selecting 493 stars (the same number as we have stars with detected planets) from the parent sample with a selection probability density $B$ that is a two-dimensional normal distribution on $T_{\rm eff}$ and $t_{\rm age}$:
\begin{equation}
    B \sim \mathcal{N}\left(\mu=(6000~\text{K}, 3~\text{Gyr}),\sigma=(500~\text{K},2.5~\text{Gyr})\right).
\end{equation}
We implement the selection via rejection sampling. 
Subsequently, we apply \autoref{algo1} to compute $\Delta P_{\rm rot}$ between our matched biased sample and the matched control. 
The histograms before and after matching for such a random sample are shown in \autoref{fig:randsamp_prematch} and \autoref{fig:randsamp_postmatch}. 
Note that the histograms in \autoref{fig:randsamp_postmatch} are wider than those in \autoref{fig:randsamp_prematch} (mostly [Fe/H] and $t_{\rm age}$) because of the bootstrapping that accounts for measurement errors.

\begin{figure}
    \centering
    \includegraphics[scale=.7]{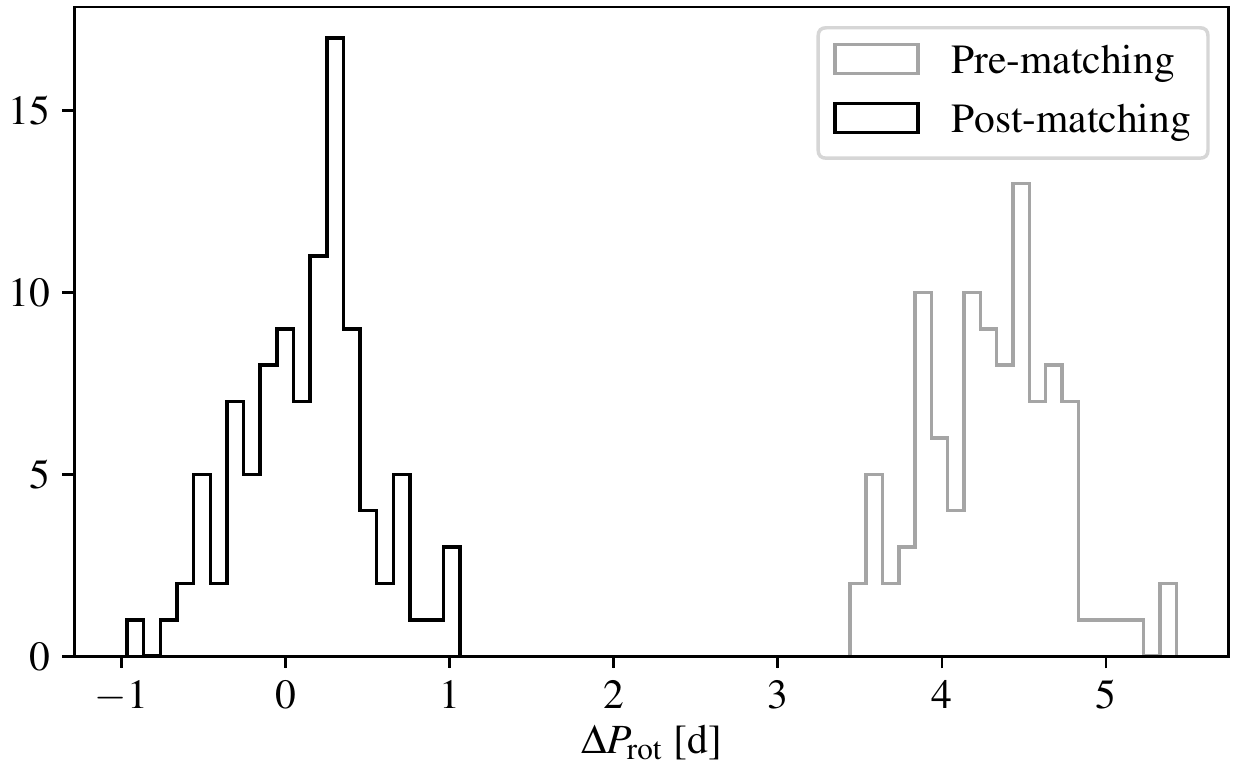}
    \caption{Histograms of the difference in stellar rotation period ($\Delta P_{\rm rot}$) between 100 realisations of a  biased random sample of stars without planets and either the (unmatched) control sample (grey) or the matched control sample (black).  
    }
    \label{fig:randsamp_distribution}
\end{figure}

The process described above (rejection sampling of 493 stars, followed by bootstrapping and matching) yields one value of $\Delta P_{\rm rot}$. 
However, given the element of randomness in creating a biased sample, we repeat the process 100 times, which results in the histogram of the 100 $\Delta P_{\rm rot}$ values shown in \autoref{fig:randsamp_distribution} with an average $\overline{\Delta P_{\rm rot, B}}=(0.14 \pm 0.33)$d. 
\par
As expected, the difference in rotation period is (on average) consistent with zero after matching. In contrast, the rotation period difference between the biased sample and the control data set before the matching is $(4.31 \pm 0.41)$d, i.e., very different from zero.

\setcounter{figure}{0} \renewcommand{\thefigure}{C.\arabic{figure}}
\setcounter{table}{0} \renewcommand{\thetable}{C.\arabic{table}}
\setcounter{algorithm}{0} \renewcommand{\thealgorithm}{C.\arabic{algorithm}}
\setcounter{equation}{0} \renewcommand{\theequation}{C.\arabic{equation}}
\renewcommand{\theHfigure}{C.\thefigure}

\section{Removing stars with radial velocity confirmation} \label{Appendix3}

\begin{figure*}
    \centering
    \includegraphics[scale=.8]{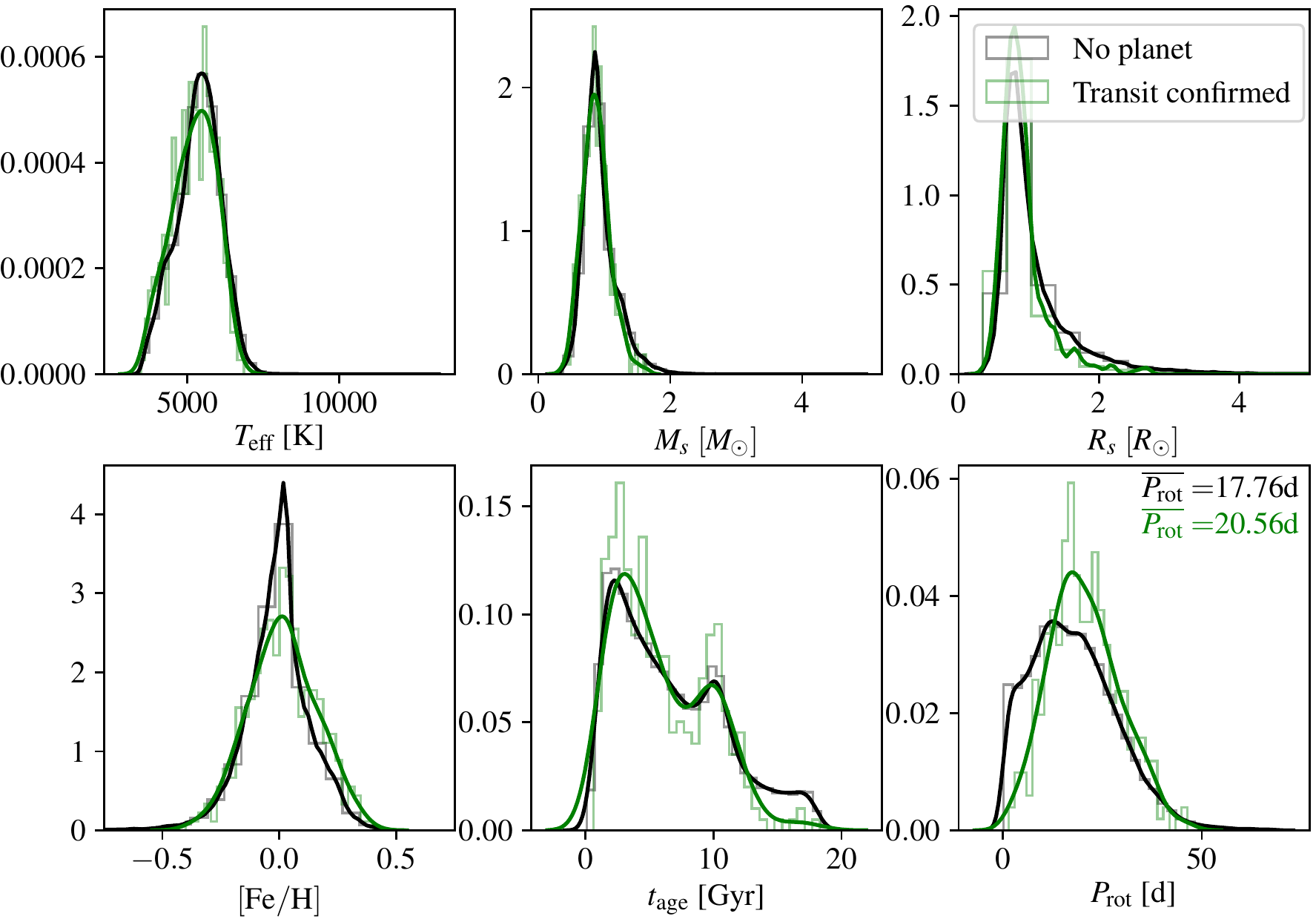}
    \caption{Histograms of the stellar properties of our control sample (shown in black) and the sample of stars with planets confirmed by TTV or statistical validation (shown in green).
    \label{fig:transitconfirmed_prematch}}
    \includegraphics[scale=.8]{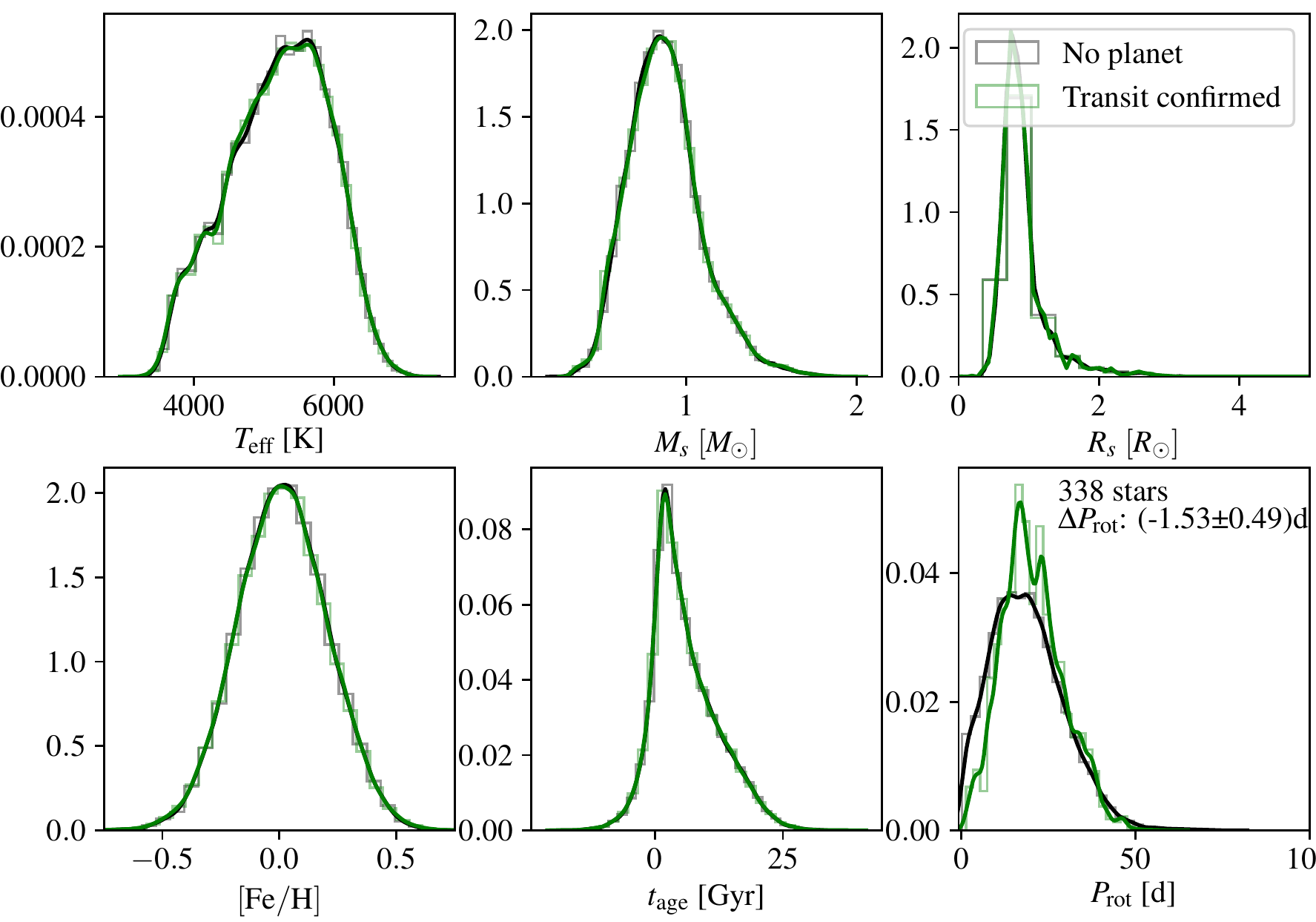}
    \caption{Histograms of the stellar properties after matching the control (shown in black) to the sample of stars with planets confirmed by TTV or statistical validation (shown in green).
    \label{fig:transitconfirmed_postmatch}}
\end{figure*}

A large fraction of the Kepler planets have been confirmed by radial velocity (RV) follow-up observations. This confirmation method introduces biases in the dataset because stellar rotation causes broadening of the Doppler profile, making the RV signal harder to detect. As a result, planets confirmed by RV should be more prevalent around slowly rotating stars. To ensure that our results are not strongly affected by this observational bias, we select in our dataset of stars with confirmed planets only those where no planet mass or semi-major axis has been measured, amounting to 338 over the original 493 stars. We manually verify the confirmation method for 40 randomly selected stars in the 338, all of which had their planet(s) confirmed either by transit-timing variation (TTV, one study for one star) or by statistical validation surveys \citep[one study for many stars, e.g.][]{Rowe2014,Morton2016}.\\
The histograms before and after matching for the `transit confirmed' dataset are shown in \autoref{fig:transitconfirmed_prematch} and \autoref{fig:transitconfirmed_postmatch}. Importantly, we can see that removing the stars with RV-confirmed planets does not change the overall result of our study: stars harbouring confirmed planets rotate statistically slower than stars without confirmed planets by $\sim$1.5 days.

\bsp	
\label{lastpage}
\end{document}